\documentclass[aps,pre, superscriptaddress,notitlepage]{revtex4-2}   

\usepackage{graphicx, xcolor}
\usepackage{hyperref}
\hypersetup{
    colorlinks = true,
    urlcolor   = blue,
    citecolor  = black,
}

\usepackage{bm}
\usepackage{comment} 
\usepackage{amsmath,amssymb, mathtools}

\providecommand*{\qmbox}[1]{\quad \mbox{#1} \quad}

\providecommand*{\qmbox}[1]{\quad \mbox{#1} \quad}
\renewcommand{\S}{\mathcal{S}}
\newcommand{\A}{A}
\renewcommand{\epsilon}{\varepsilon}

\newcommand{\binfV}{\bm{\mathcal{V}}}
\newcommand{\binfE}{\bm{\mathcal{E}}}
\newcommand{\binfVhat}{\hat{\bm{\mathcal{V}}}}
\newcommand{\binfEhat}{\hat{\bm{\mathcal{E}}}}
\newcommand{\bv}{\bm{v}}
\newcommand{\bx}{\bm{x}}
\newcommand{\br}{\bm{r}}
\newcommand{\bV}{\bm{V}}

\newcommand{\bsigma}{\bm{\sigma}}
\newcommand{\avg}[1]{\left<#1\right>}

\newcommand{\inteS}{\int_{S^{(1)}}}

\newcommand{\myquad}[1][1]{\hspace*{#1em}\ignorespaces}
\newcommand{\dS}{\mathrm{d}S}
\newcommand{\dV}{\mathrm{d}V}

\begin{document}

\title{Inertio-viscous interactions between particles in oscillatory flow}

\author{Xiaokang Zhang}
\affiliation{Department of Mechanical Engineering,
University of California, Riverside, CA 92521}
\author{Bhargav Rallabandi}
\affiliation{Department of Mechanical Engineering,
University of California, Riverside, CA 92521}
\email{bhargav@engr.ucr.edu}

\begin{abstract}
We study the interaction between a pair of particles suspended in a uniform oscillatory flow. The time-averaged behavior of particles under these conditions, driven by inertial and viscous effects, is explored through a theoretical framework relying on small oscillation amplitude. We approximate the  oscillatory flow in terms of dual multipole expansions, with which we compute time-averaged interaction forces using the Lorentz reciprocal theorem. The results demonstrate excellent agreement with existing numerical data. We then develop analytic approximations for the force, valid when Stokes layers surrounding the particles do not overlap. Finally, we show how the same formalism can be generalized to the situation where the particles are free to oscillate and drift in response to the applied flow. The theory thus provides an efficient means to calculate nonlinear particle interactions in oscillatory flows. 
\end{abstract}

\maketitle


\section{Introduction} \label{sec:level1}
Particles in oscillatory flows and acoustic fields exhibit complex behavior due to advective nonlinearities. Due to inertial and viscous effects, fluid oscillations around particles lead to time-averaged flows, known as streaming \cite{riley2001steady}. When the flow is spatially nonuniform, particles may also experience time-averaged secondary radiation forces \cite{king1934acoustic, Gorkov1962, settnes2012forces}. The combination of these effects can lead to a steady drift of suspended particles over many oscillation cycles. 

These phenomena have been used for different applications, such as particle sorting \cite{friend2011microscale} and focusing in microfludics \cite{mutlu2018oscillatory,rufo2022acoustofluidics,yang2022harmonic}. It is also useful in the levitation of particles and droplets \cite{foresti2014acoustophoretic, andrade2020acoustic}.  The levitation of particles also makes it easier to observe phenomena which are hard to study due to gravity; for example, \cite{lee2018collisional} use acoustic levitation to study electrostatic charging of fine particles due to collisions. Radiation forces have been used in the non-contact extraction and manipulation of droplets from liquid-liquid interfaces \cite{lirette2019ultrasonic}. 

Many examples involve multiple particles, and interactions between these particles may play an important role in the observed behaviors. 
Experimental studies show that suspensions of particles exposed to oscillations demonstrate collective motion such as the formation of chains \cite{Klotsa_2009}, lattices (long-range attraction and short-range repulsion) \cite{voth2002ordered}, or clusters \cite{sazhin2008particle,lim2019cluster}. Developing a predictive quantitative understanding of these behaviors is challenging, as multiple time and length scales are involved, the phenomena rely on nonlinear inertial effects. Recent computational work on the topic has focused on understanding the interactions between a pair of particles. One approach is to develop a perturbation expansion in small oscillation amplitudes. Although the reduced system of equations still typically requires numerical solutions \cite{ingber2013particle,fabre2017acoustic}, it yields insight into scaling behaviors and symmetries of the time-averaged dynamics. Direct numerical simulations (DNS) resolving all length and time scales have yielded insight into particle interactions \cite{klotsa2007interaction, Kleischmann_Luzzatto-Fegiz_Meiburg_Vowinckel_2024}, but can be computationally expensive.  


The focus of the present article is to gain analytic insight into to the interactions of two particles in oscillatory flow, filling the gap between past experimental and computational studies. We build on recent progress on the dynamics of single particles in oscillatory flows, accounting for both inertial and viscous effects \cite{zhang2014acoustic,agarwal2021unrecognized, agarwal2024density, zhang2024particle}. We focus on the regime of small oscillation amplitudes, which allows us to split the problem into fast (oscillatory) and slow (time-averaged) scales (section \ref{SecSmallAmp}). Using this decomposition, we develop a theory that combines multipole expansions for the oscillatory flow and the Lorentz reciprocal theorem  (sections \ref{SecLRT} and \ref{SecDualMult}). Together, these techniques lead to a highly efficient semi-analytic framework for the time-averaged interaction forces (section \ref{SecForce}), which we find to be in good agreement with numerical solutions of the small-amplitude scheme \cite{fabre2017acoustic}.  Our framework yields fully analytic results in the limit where the particles are sufficiently far apart for their Stokes boundary layers to not overlap, providing quantitative insights into the scaling of forces with the inter-particle distance and oscillatory Stokes number. We then extend the framework to calculate the time-averaged drift velocity of freely suspended particles, finding good agreement with DNS (section \ref{SecFreelySuspended}). We discuss possible extensions of the framework before concluding in section \ref{SecConc}. 




\section{Problem Setup}  
\begin{figure}
    \centering
    \includegraphics[width=0.70\textwidth]{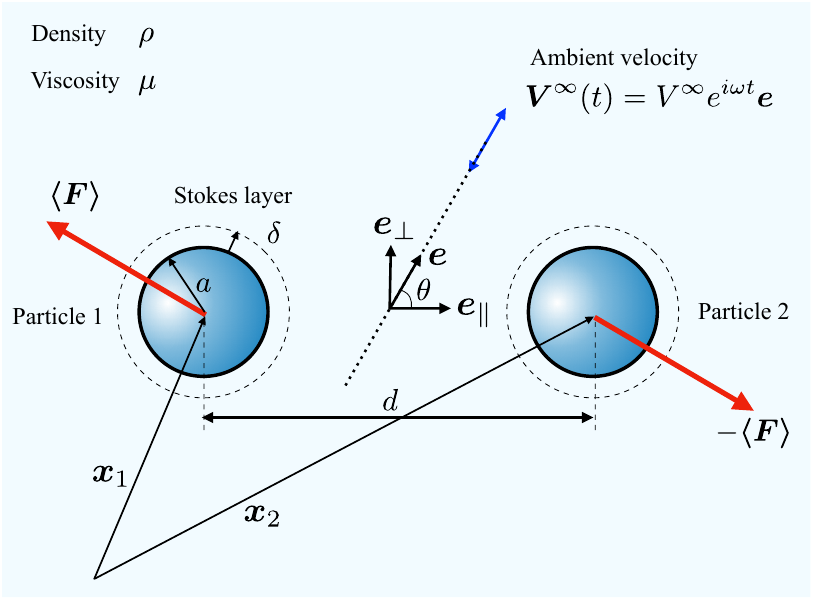}
    \caption{Sketch showing two identical particles suspended in a uniform ambient oscillatory flow $\bm{V}^{\infty}(t)$. Hydrodynamic interactions between the particles, due to advective nonlinearities, drive secondary time-averaged forces and drift velocities of the particles.}
    \label{fig.Setup}
\end{figure}

We consider a pair of identical spherical particles of radius $a$, suspended in a spatially uniform oscillatory flow $\bV^{\infty}(t) = V^{\infty} \bm{e}\cos\left(\omega t\right)$, where $V^{\infty}$ is the characteristic speed of the flow, $\bm{e}$ is the unit vector along the axis of flow, and $\omega$ is the angular frequency of oscillation. The centers of the particles separated by a distance $d > 2a$  and lie on an axis defined by the unit vector $\bm{e}_{\parallel}$ pointing from particle 1 to particle 2, and oriented at an angle $\theta$ relative to the flow axis $\bm{e}$ (figure \ref{fig.Setup}). We also define a unit vector $\bm{e}_{\perp}$ perpendicular to the particle axis in the $\bm{e}$-$\bm{e}_{\parallel}$ plane. The presence of the particles leads to oscillatory flow gradients, which in turn produce secondary ``streaming'' flows with nonzero time average due to inertial effects. These inertial flows exert time-averaged forces on the particles, which are the quantities of interest here. 

In most practical settings, such a pair of particles would be free to oscillate in response to the flow. In this case, the time-averaged inertial forces are balanced by viscous drag, leading to a drift of the particles over many oscillations. To understand the practically important case of mobile particles, it is convenient to first consider particles that are held stationary. We will show later that the case of mobile particles can be understood through a change in reference frame as long as the particles have identical properties.

We decompose the flow around the particles $\bm{v}(\bm{x},t)$ into an ambient contribution $\bm{V}^\infty(t)$ and a disturbance contribution $\bv^{d}(\bx,t)$, so that $\bv(\bm{x},t) = \bV^{\infty}(t) + \bv^{d}(\bm{x},t)$.  The amplitude of the flow oscillation relative to the particle radius is $\varepsilon = V^{\infty}/(a \omega)$. The disturbance velocity is characterized by a Stokes Number $\mathcal{S} = \omega a^2/\nu$, which is the measure of the ratio of inertial to viscous forces over an oscillation period. The vorticity of the oscillatory flow is confined to Stokes layers of dimensionless thickness $\delta = \sqrt{2/\mathcal{S}}$ (defined in units of particle radius) around each particle, where inertial and viscous effects are comparable. 

We use the particle radius $a$ as a characteristic length, the inverse angular frequency $\omega^{-1}$ as characteristic time, and the ambient flow speed $V^{\infty} = \varepsilon a \omega$ as the characteristic velocity. The flow is the governed by the dimensionless Navier-Stokes equations
\begin{subequations} \label{NS}
    \begin{align}
        &{\S}\left(\frac{\partial\bv}{\partial{t}}+\varepsilon\bv\cdot\nabla\bv\right)=\nabla\cdot\bsigma, \quad \nabla\cdot\bv=0, \qmbox{with}\\
    &\bv|_{S^{(1)}}=\bv|_{S^{(2)}} = \bm{0},\\
    &\bm{v}(|\bm{x}| \rightarrow\infty, t) = \bm{e} \cos t =  (\bm{e}_{\parallel} \sin \theta + \bm{e}_{\perp} \cos \theta) \cos t,
    \end{align}
\end{subequations}
where $S^{(1)}$ and $S^{(2)}$ represent the surfaces of particles $1$ and $2$, respectively, and $\bsigma=-p\bm{I}+\left(\nabla\bv+\nabla\bv^\mathsf{T}\right)$ is the dimensionless Newtonian stress tensor (rescaled with $\mu V^{\infty}/a$). The goal of the present study is to calculate the time-averaged interaction forces between particles held stationary, and later, time-averaged drift velocities of freely suspended particles. 
\section{Small amplitude theory} \label{SecSmallAmp}
The full nonlinear flow is complicated and requires numerical methods to resolve. In typical applications \cite{Klotsa_2009,voth2002ordered}, the dimensionless amplitude of oscillation $\epsilon$ is small. To gain analytic insight in this limit, we perform a small-amplitude expansion with the parameter $\epsilon$,
\begin{align}
    (\bm{v}, \bm{\sigma}) &= (\bm{v}_1, \bm{\sigma}_1) + \epsilon (\bm{v}_2, \bm{\sigma}_2)+ O(\epsilon^2),
\end{align}
The primary flow $\bv_1(\bx, t)$ oscillates with a frequency $\omega$ and scales with the applied ambient flow. The secondary flow $\bv_2(\bx, t)$ involves a combination of a time-dependent component (with frequency 2$\omega$) and a steady (or time-averaged) component. We are ultimately interested in the time-averaged flow, since it is associated with a time-averaged force on the particles. Here and below, we use subscripts to indicate orders of $\epsilon$, and superscripts to identify particles. 

Substituting the above expansion into \eqref{NS} and separating orders of $\epsilon$, we obtain the governing equations for the primary and secondary flows. The primary flow satisfies
\begin{subequations}\label{PrimaryGE}
  \begin{align} 
   & \S\frac{\partial\bv_1}{\partial t}=\nabla\cdot\bsigma_1,\quad\nabla\cdot\bv_1=0, \qmbox{with} \\
   &\bm{v}_1|_{S^{(1)}} = \bm{v}_1|_{S^{(2)}}=\bm{0},\quad  \bm{v}_1(|\bm{x}| \rightarrow\infty, t) = (\bm{e}_{\parallel} \sin \theta + \bm{e}_{\perp} \cos \theta) \cos t. \label{eqn.1stGE}
\end{align}  
\end{subequations}
Averaging \eqref{NS} over an oscillation cycle, we see that the secondary flow is governed by
\begin{subequations} \label{SecondaryGE}
 \begin{align}
    &{\nabla}\cdot\avg{{\bsigma}_2-\S{\bv}_1{\bv}_1}=\bm{0}
      ,\quad \nabla\cdot\avg{\bv_2} = 0, \qmbox{with} \\
    &\avg{\bm{v}_2}|_{S^{(1)}} =  \avg{\bm{v}_2}|_{S^{(2)}} =  \avg{\bm{v}_2}(|\bm{x}| \rightarrow\infty) = \bm{0},\label{eqn.2ndGE}
 \end{align}
\end{subequations}
where angle brackets define a time-average over an oscillation according to $\left<g\right>(\bm{x}) = (2 \pi)^{-1} \int_{0}^{2\pi} g(\bm{x}, t) dt$. 

We are ultimately interested in the time-averaged force on any one of the particles.  Since there is no net force on the system, the time-averaged force on each particle is identical in magnitude but opposite in direction to the other particle. Using the formulation of \cite{doinikov1994acoustic}, the time-averaged force on particle 1, non-dimensionalized with $\mu a \epsilon V^{\infty}$, is 
\begin{align} \label{eqn.doi}
    \left<\bm{F}\right> = \left<\bm{F}\right>^{(1)} = -\left<\bm{F}\right>^{(2)}= \inteS\bm{ n}\cdot\left<\bsigma_2-\S\bv_1\bv_1\right>\,\dS.
\end{align}
The ambient flow is uniform and oscillatory, and thus makes no contribution to the force. Consequently, we replace the kernel of \eqref{eqn.doi} by the disturbance stress $\left<\bsigma_2-\S\bv_1\bv_1\right>^d$, where $f^d = f - f^\infty$ denotes the disturbance of any flow quantity $f$, with $f^\infty$ representing that quantity in the absence of the particles.


\section{Reciprocal theorem for the time-averaged force} \label{SecLRT}

The integrand in \eqref{eqn.doi} involves the secondary time-averaged stress $\avg{\bm{\sigma}_2}$ as well as the Reynolds stress due to the primary flow. Even in the perturbation framework, the linear equations \eqref{PrimaryGE} and \eqref{SecondaryGE} must be solved numerically (this was the approach adopted by \cite{fabre2017acoustic}), partly due to the complicated two-sphere geometry, and partly due to the potentially large separation of length scales between $\delta$, $a$ and $d$. While the primary flow yields to analytic approximation (Sec. \ref{SecDualMult}), the secondary flow problem remains analytically intractable as it involves a body force that inherits the structure of $\bm{v}_1$. 

In order to gain insight into the time-averaged force without solving the secondary problem, we utilize the  Lorentz reciprocal theorem \cite{stone1996propulsion, masoud2019reciprocal}. We introduce an auxiliary flow $(\hat{\bm{v}},\hat{\bm{\sigma}})$, which we define as the steady Stokes flow around two spheres translating with velocity $\pm\hat{\bm{V}}$, thereby satisfying
\begin{align} \label{eqn.AuxGE}
    \nabla\cdot\hat{\bsigma}=\bm{0},\quad  \nabla\cdot\hat{\bv}=0,\qquad \text{subject to}\qquad  \hat{\bv}|_{S^{(1)}}= -\hat{\bv}|_{S^{(2)}}=\hat{\bV},\quad\hat{\bv}|_{r\to\infty}=\bm{0}.
\end{align} 

Combining the governing equations for the secondary time-averaged flow \eqref{SecondaryGE} (as noted earlier, we only consider the disturbance contribution) and the auxiliary flow \eqref{eqn.AuxGE}, we construct a symmetry relation \cite{zhang2024particle}
\begin{align}
    \nabla \cdot \left<\bsigma_2 - \S\bv_1 \bv_1 \right>^d\cdot \hat{\bv} = \nabla \cdot \hat{\bsigma} \cdot \left<\bv_2\right>^d,
\end{align}
which we then recast as
\begin{align}
    \nabla \cdot \left[\left<\bsigma_2 - \S\bv_1 \bv_1 \right>^d\cdot \hat{\bv} \right]-\left<\bsigma_2 - \S\bv_1 \bv_1 \right>^d:\nabla\hat{\bv}= \nabla \cdot \left(\hat{\bsigma} \cdot \left<\bv_2\right>^d\right)-\hat{\bsigma}:\nabla\left<{\bv}_2\right>^d.
\end{align}
We integrate over the fluid volume $V_f$ and apply Gauss's theorem to obtain

\begin{align}  \label{RTItermediate}
    \begin{split}
&\int_{S^{(1)}+S^{(2)} + S^{\infty}}\bm{n}\cdot\left<\bsigma_2 - \S\bv_1 \bv_1\right>^d\cdot\bm{\hat{v}}\dS - \int_{V_f}\left<\S\bv_1 \bv_1\right>^d:\nabla{\hat{\bv}}~\dV\\&\myquad[5] =\int_{S^{(1)}+S^{(2)} + S^{\infty}}\bm{n}\cdot\hat{\bsigma}\cdot \left<\bm{v}_2\right>^d\dS,
    \end{split}
\end{align} 
where $\bm{n}$ is the fluid-facing normal vector. 

The integrals at the bounding surface at infinity, $S^{\infty}$, vanish since the secondary fields only contain disturbance quantities that decay rapidly in the far field. Noting that the auxiliary velocity is constant on the particle surfaces and using \eqref{eqn.doi}, the first term on the left-hand side of \eqref{RTItermediate} simplifies to $\avg{\bm{F}}^{(1)}\cdot\hat{\bV} + \avg{\bm{F}}^{(2)}\cdot(-\hat{\bV}) = 2\avg{\bm{F}}\cdot \hat{\bV}$. Similarly, the term on the right simplifies to $2\hat{\bm{F}}\cdot\avg{\bm{V}_2}$, where $\hat{\bm{F}} = \int_{S^{(1)}} \bm{n} \cdot \hat{\bm{\sigma}}~\dS $ is the force on particle 1 due to the  auxiliary flow (the auxiliary force on particle 2 is equal and opposite), and $\avg{\bm{V}_2}$ is the time-averaged velocity of particle 1. We note that $\avg{\bm{V}_2} = \bm{0}$ for stationary particles, but we retain it in the interest of generality (later we will relax the assumption of stationary particles). Thus, we obtain
\begin{align} \label{RT1}
     \avg{\bm{F}}\cdot\hat{\bV} - \hat{\bm{F}}\cdot\avg{\bm{V}_2} = \frac{\S}{2} \int_{V_f}\left<\bv_1 \bv_1\right>^d:\nabla{\hat{\bv}}~\dV.
\end{align}
The above result allows for the computation of the secondary force on either particle along an arbitrary direction $\hat{\bm{V}}$, without requiring a solution to the secondary flow. However, the integrand in \eqref{RT1} decays slowly at large distances from the particles, making numerical evaluation (and later, analytic approximations) inconvenient. We therefore recast this integrand in terms of the primary vorticity $\bm{\omega}_1 = \nabla \times \bm{v}_1$, which decays exponentially away from the particles on length scales of $O(\delta)$. To this end, we write the integrand in \eqref{RT1} according to the identity ${\bv_1\bv_1:\nabla{\hat{\bv}} =\nabla\cdot\left(\bv_1\bv_1 \cdot\hat{\bv}\right)-\left(\bv_1\cdot\nabla\bv_1\right)\cdot\hat{\bv}-\bv_1\cdot\hat{\bv}\left(\nabla\cdot\bv_1\right)}$, 
where the last term is zero due to incompressibility. We then use the identity $\bv_1\cdot\nabla\bv_1=\frac{1}{2}\nabla|\bv_1^2|- (\bv_1\times \bm{\omega}_1)$. Substituting these relations into \eqref{RT1} and using Gauss's theorem, we obtain
\begin{align}
    \begin{split}
    &\avg{\bm{F}}\cdot\hat{\bV} - \hat{\bm{F}}\cdot\avg{\bm{V}_2} \\&\myquad[3] = \frac{\S}{2} \left[\int_{V_f} \left<\bv_1 \times \bm{\omega}_1 \right>^d\cdot{\hat{\bv}}~\dV - \int_{S^{(1)} + S^{(2)}} \bm{n} \cdot \left<\bv_1 \bv_1 - \frac{|\bm{v}_1|^2}{2} \bm{I}  \right>^d\cdot\hat{\bv} ~ \dS\right].
    \end{split}
\end{align}
The surface integrals go to zero due to the no-slip condition on the particle surfaces, leading to 
\begin{align} \label{RT2}
    \avg{\bm{F}}\cdot\hat{\bV} - \hat{\bm{F}}\cdot\avg{\bm{V}_2} =\frac{\S}{2}\int_{V_f} \left(\bv_1\times \bm{\omega}_1\right)\cdot\bm{\hat{v}}~ \dV.
\end{align}
Equation \eqref{RT2} is an exact writing of \eqref{eqn.doi}. It directly involves the primary vorticity, which is confined to near-surface regions around the particles (Stokes boundary layers) of thickness $\delta$. Outside these regions, the vorticity -- and thus the integrand -- is exponentially small, making \eqref{RT2} much more computationally efficient than \eqref{RT1}. Later we will see that the ``localization'' of the integrand is amenable to analytic approximations.

\section{Primary and auxiliary flows: Dual multipole expansions} \label{SecDualMult}

We have shown that the only information needed to calculate the time-averaged force are the primary and auxiliary flow. Since exact solutions to these problems are not generally possible in two-sphere geometries, we develop approximations using dual multipole expansions. 

\subsection{Primary flow}
It is useful to represent all primary flow quantities as complex phasors proportional to $e^{it}$, the real parts of which represent the solution to \eqref{PrimaryGE}. We write the primary velocity as a superposition of the ambient and disturbance flows created by each particle $\bv_1= \bV_1^{\infty}+\bv_1^{d1}+\bv_1^{d2}$.
We approximate the disturbance flow of each particle by a truncated multipole expansion, retaining terms corresponding to translating and straining modes of flow. 

Consider a stationary particle $i$ suspended in a linear oscillatory flow field $(\binfV^i + \bm{r}_i \cdot \binfE^i) e^{it}$, where $\bm{r} = \bm{x} - \bm{x}_i$ is the position vector relative to the center of particle $i$, and $\binfV^{i}$ and $\binfE^{i}$ are the ``effective'' velocity and rate-of-strain (symmetric and traceless) felt locally by particle $i$. Then, the disturbance flow around particle $i$ has the general structure \cite{zhang2024particle,agarwal2024density}
\begin{align}\label{eqn.dist1st}
   \bv_1^{di}(\bm{x},t) = \left(\bm{D}(\br_i,\lambda)\cdot\binfV^{i} + \bm{Q}(\br_i,\lambda):\binfE^{i} + \dots \right)e^{it},
\end{align}
where $\lambda = \sqrt{i \S} = (1+i)/\delta$. The dipole $\bm{D}$ and quadrupole $\bm{Q}$ tensor fields are solutions of \eqref{PrimaryGE}, given by \cite{zhang2024particle}

\begin{subequations}
    \begin{align}\label{eqn.DQ}
        \begin{split}
    \bm{D}(\br,\lambda)&=d_1\left[\frac{\bm{I}}{r^3}-3\frac{\br\br}{r^5}\right]\\&\myquad[1]+d_2\left[e^{-R}\left(\frac{1}{R}+\frac{1}{R^2}+\frac{1}{R^3}\right)\bm{I}-e^{-R}\left(\frac{1}{R}+\frac{3}{R^2}+\frac{3}{R^3}\right)\frac{\br\br}{r^2}\right],
        \end{split}
        \\
        \begin{split}
    \bm{Q}(\br,\lambda)&=q_1\left[6\frac{\bm{I}\br}{r^5}-15\frac{\br\br\br}{r^7}\right]\\&\myquad[1]+q_2\left[e^{-R}\left(1+\frac{3}{R}+\frac{6}{R^2}+\frac{6}{R^3}\right)\frac{\bm{I}\br}{r^2}-e^{-R}\left(1+\frac{6}{R}+\frac{15}{R^2}+\frac{15}{R^3}\right)\frac{\br\br\br}{r^4}\right],
\end{split}
    \end{align}
\end{subequations} 
with coefficients 
\begin{align}
    d_1=-\frac{3+3\lambda+\lambda^2}{2\lambda^2}, \myquad
    d_2=\frac{3e^{\lambda}\lambda}{2}, \myquad
    q_1=\frac{15+15\lambda+6\lambda^2+\lambda^3}{9\lambda^2(1+\lambda)},  \myquad
    q_2=-\frac{5e^{\lambda}\lambda}{3(1+\lambda)}.
\end{align}

Next, we recognize that the effective quantities $\binfV^i$ and $\binfE^i$ result from a combination of the background flow $\bm{V}^{\infty}$ and the disturbance created by the other particle. Neglecting terms involving the curvature of the disturbance fields, we obtain the relations
\begin{subequations}
    \begin{align}\label{eqn.Effective}
        \binfV^{1}&=\bV_1^{\infty}+\bv_1^{d2}(\bm{x}_1)\\
        \binfV^{2}&=\bV_1^{\infty}+\bv_1^{d1}(\bm{x}_2)\\
        \binfE^{1}&=\overbracket{\nabla\bv_1^{d2}}(\bm{x}_1)\\
        \binfE^{2}&=\overbracket{\nabla\bv_1^{d1}}(\bm{x}_2),
    \end{align}
\end{subequations}
where $\overbracket{\bm{T}}$ represents the symmetric and traceless part of a rank-2 tensor $\bm{T}$. Equation \eqref{eqn.Effective} is a linear system of equations for the effective quantities $\binfV^i$ and $\binfE^i$, and can be solved either analytically or numerically for any configuration of the particles, yielding an approximation to  the primary flow. We note that we have neglected contributions due to the antisymmetric part of the velocity gradient tensor, since these terms decay exponentially away from the particles on length scales of $\delta$.  We use \eqref{eqn.dist1st} to calculate the primary vorticity $\bm{\omega}_1$. 

\subsection{Auxiliary flow multipole expansion}
\begin{figure}
    \centering
    \includegraphics[width=0.6\textwidth]{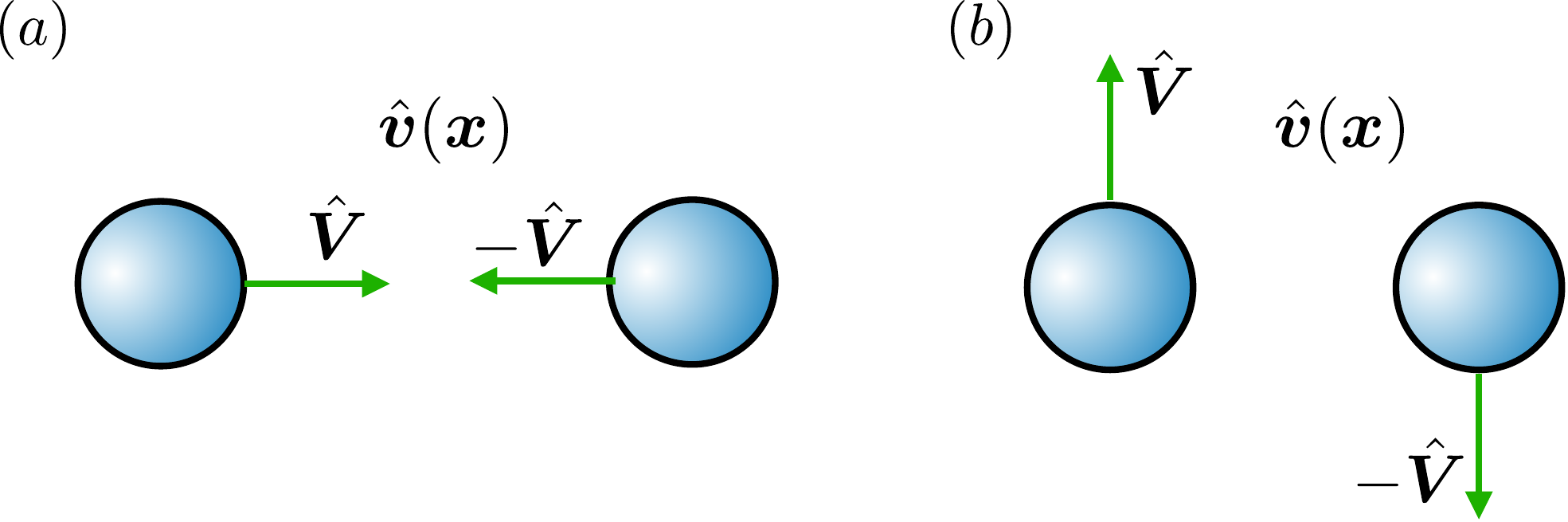}
    \caption{Auxiliary flow: particles translating with opposite velocities $\pm \hat{\bm{V}}$. (a) Particles moving towards each other. (b) Particles moving normal to their line-of-centers.}
    \label{fig.aux}
\end{figure}

The auxiliary flow is constructed similarly to the primary flow using a multipole expansion. We decompose the auxiliary velocity as a superposition of  disturbance flows created by each particle $\hat{\bv}= \hat{\bv}^{(1)}+\hat{\bv}^{(2)}$. A particle $i$ translating with velocity $\hat{\bm{V}}^i$ exposed to a linear flow $(\hat{\binfV}^i + \bm{r}_i \cdot \hat{\binfE}^i)$ produces a disturbance flow 
\begin{align}\label{eqn.aux1st}
    \hat{\bv}^{i}= \hat{\bm{D}}(\br)\cdot\left(\hat{\bV}^{i}-\hat{\binfV}^{i}\right)+ \hat{\bm{Q}}(\br):\hat{\binfE}^{i} + \dots,
\end{align}
where the tensor fields $\hat{\bm{D}}(\br)$ and $\hat{\bm{Q}}(\br)$ are
\begin{subequations}
    \begin{align}
    \hat{\bm{D}}(\br) &= \frac{3}{4} \left[ \left( \frac{\bm{I}}{r} + \frac{\bm{r} \bm{r}}{r^3} \right) + \frac{1}{3} \left( \frac{\bm{I}}{r^3} - \frac{3 \bm{r} \bm{r}}{r^5}  \right)  \right], \\
    \hat{\bm{Q}}(\br) &= -\frac{5}{2}\left[\frac{\br \br}{r^5} + \frac{1}{5} \left( \frac{\bm{I}}{r^5} - 5 \frac{\bm{r} \bm{r}}{r^7} \right) \right] \bm{r}.
    \end{align}
\end{subequations}
The effective velocity $\binfVhat^i$ and rate of strain $\binfEhat^i$ are due to the disturbance created from the other particle evaluated at the center of particle $i$. We also include Fax\'{e}n corrections as is standard in particle hydrodynamics in Stokes flows, making use of standard results for spheres \cite{batchelor1972hydrodynamic}. Thus, the effective velocities and rates of strain are governed by \cite{brady1988stokesian,durlofsky1987dynamic}
\begin{subequations}\label{eqn.syseqnaux}
    \begin{align}
        \binfVhat^{1}&=\left(1+\frac{1}{6}\nabla^2\right)\hat{{\bm{v}}}^{d2}(\bm{x}_1)\\
        \binfVhat^{2}&=\left(1+\frac{1}{6}\nabla^2\right)\hat{{\bm{v}}}^{d1}(\bm{x}_2)\\
        \binfEhat^{1}&=\left(1+\frac{1}{10}\nabla^2\right)\overbracket{\nabla\hat{{\bm{v}}}^{d2}}(\bm{x}_1)\\
        \binfEhat^{2}&=\left(1+\frac{1}{10}\nabla^2\right)\overbracket{\nabla\hat{{\bm{v}}}^{d1}}(\bm{x}_2).
    \end{align}
\end{subequations}
It is useful to note that $-6 \pi \mu a \left(\hat{\bV}^{i}-\hat{\binfV}^{i}\right)$ and $(20 \pi/3)  \mu a^3 \binfEhat^i$ are, respectively, the force and stresslet exerted by the auxiliary flow on the particle $i$. Similarly to the primary flow, we solve the system of equations \eqref{eqn.syseqnaux} to calculate the effective quantities $\binfVhat^i$ and $\binfEhat^i$, thereby determining the auxiliary flow. There are two auxiliary flow configurations of interest: particles moving towards each other, and particles moving normal to their connecting axis in opposite directions (figure \ref{fig.aux}). These configurations, respectively, let us calculate components of $\avg{\bm{F}}$ in the $\bm{e}_{\parallel}$ and $\bm{e}_{\perp}$ directions. 

\section{Time-averaged forces on stationary particles} \label{SecForce}

\subsection{Calculation of the force}
We first consider fixed particles ($\avg{\bm{V}_2} = \bm{0}$) and compute the time-averaged forces. In Section \ref{SecFreelySuspended} we calculate time-averaged drift velocities of freely suspended particles. Both primary and auxiliary flows are now known up to the level of force and mass dipoles and depend on the parameters $\S$, $D = d/a$, and $\theta$. We substitute these fields into the expression \eqref{RT2}, where all quantities in the integrand are now analytically known. We execute the volume integral numerically since the fluid volume has a somewhat complicated geometry. However, since the primary vorticity decays exponentially outside Stokes layers, the domain of integration only needs to be a few multiples of the Stokes layer thickness $\delta$ to achieve convergence. This procedure yields the projection of $\avg{\bm{F}}$ along an arbitrarily chosen direction  $\hat{\bm{V}}$.  

To identify the general structure of the force, we appeal to tensor symmetries. The force depends on the ambient velocity $\bV^{\infty}$ (which in dimensionless coordinates is simply $\bm{e}$), the unit vector $\bm{e}_{\parallel}$ connecting the particles, the Stokes Number $\mathcal{S}$, and the distance $D$. Furthermore, the force is quadratic in oscillatory velocity, and is therefore quadratic in  $\bm{e}$. The most general expression for a force satisfying these properties is  (restoring dimensions and using an inertial scale)
\begin{equation} \label{eqn.generalF}
\avg{\bm{F}} = \rho (V^{\infty})^2 a^2 \left[\alpha\left(\bm{e}\cdot\bm{e}\right)\bm{e}_{\parallel}+\beta(\bm{e}\cdot\bm{e}_{\parallel})^2\bm{e}_{\parallel}+\gamma(\bm{e}\cdot\bm{e}_{\parallel})\bm{e}\right],
\end{equation}
where $\alpha$, $\beta$ and $\gamma$ are scalar functions of $D$ and $\S$. We substitute $\bm{e} = \bm{e}_{\parallel} \cos \theta + \bm{e}_{\perp} \sin \theta$ to recast the above expression in terms of $\theta$, finding
\begin{align}\label{eqn.Fabre}
    \avg{\bm{F}}= \rho (V^{\infty})^2 a^2\left[F_{AA} \cos^2\theta \, \bm{e}_{\parallel} + F_{TT} \sin^2\theta\, \bm{e}_{\parallel} + F_{AT}\sin{\theta}\cos\theta\, \bm{e}_{\perp}\right],
\end{align}
where  $F_{AA}=\left(\alpha+\beta+\gamma\right)$, $F_{TT} = \alpha$, and $F_{AT} = \gamma$ (all functions of $\S$ and $D$). Equations \eqref{eqn.generalF} and \eqref{eqn.Fabre}, obtained here through straightforward tensorial symmetries, is identical to the result obtained by \cite{fabre2017acoustic} through more detailed arguments involving the spatial structure of the secondary flow field.  The $AA$ contribution is quadratic in the axial component of the ambient flow ($\cos \theta$), the $TT$ contribution is quadratic in the transverse velocity component ($\sin \theta$), while the $AT$ contribution involves a product of axial and transverse velocity components.  We observe that $F_{AA}$ and $F_{TT}$ contributions are both associated with forces along the axis connecting the particle (positive values indicate attraction), while the $F_{AT}$ term is associated with ``reorienting'' forces transverse to the connecting axis. 

To extract the $F$ coefficients from \eqref{RT2}, we fix the particle orientation and make specific choices of $\theta$ (which sets the ambient flow orientation $\bm{e}$) and the auxiliary velocity $\hat{\bm{V}}$. For example, picking $\theta = 0$ and $\hat{\bm{V}} = \bm{e}_{\parallel}$ and comparing with \eqref{eqn.Fabre} leads to $F_{AA}$. Choosing  $\theta = \pi/2$ for the same auxiliary flow yields $F_{TT}$, whereas picking $\theta = \pi/4$ and $\hat{\bm{V}} = \bm{e}_{\perp}$ determines  $F_{AT}$.

\begin{figure}
    \centering
    \includegraphics[width=1.0\textwidth]{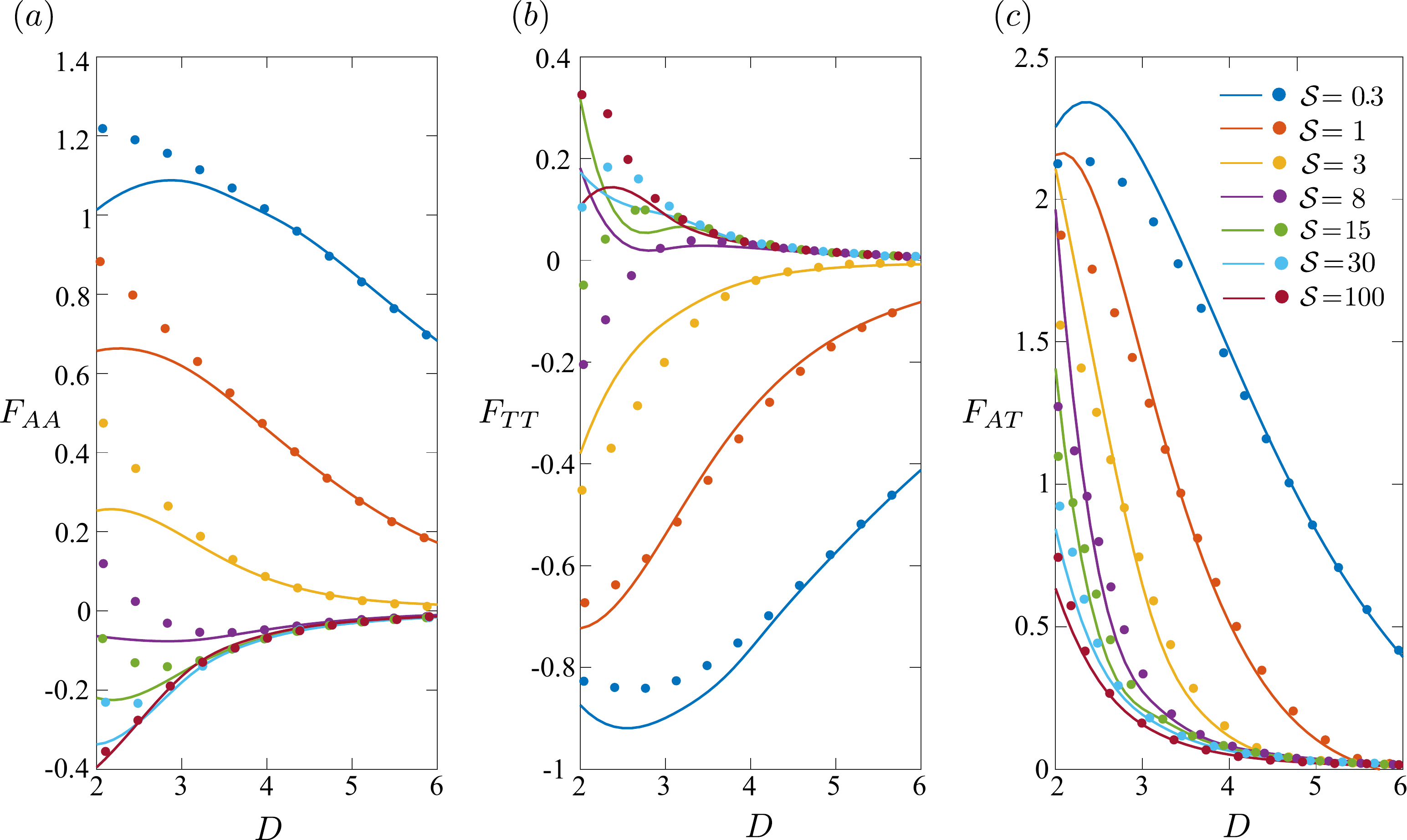}
    \caption{Comparison of time-averaged forces in all configurations --- (a) axial ($AA$), (b) transverse ($AT$), (c) reorienting ($AT$) ---  showing the numerical calculations of \cite{fabre2017acoustic} (symbols) and the present semi-analytic theory (solid) for different distances $D$ and Stokes numbers $\S$. Each curve corresponds to a single value of $\S$; curves and circles which share the same color are of the same value of $\S$.}
    \label{FabreCompare}
\end{figure}

With these three combinations we identify all three $F$ coefficients for any $D$ and $\S$.  We plot them in figure \ref{FabreCompare} against $D$ for different $\S$. The results of our semi-analytic theory are shown as solid curves, while the numerical calculations of \cite{fabre2017acoustic}, which involve numerical solutions of both primary and  secondary flows, are indicated as symbols. The present results are in very good agreement with the numerical solutions of \cite{fabre2017acoustic} for all separation distances, despite the fact that we truncated the multipole expansion (formally accurate at large $D$) at just two terms for both the primary and auxiliary flows. The two-term theory is essentially indistinguishable from the numerical solutions for $D \gtrsim 4$ (corresponding to one diameter of surface separation), and the agreement remains reasonably accurate down to contact ($D = 2$). 

In figure \ref{FabreCompare}(a, b), positive values represent attractive forces while negative values represent repulsion. As we can see from the plots, the direction of the force is determined by a combination of distance, Stokes number and configuration. In general, the  magnitude of the force drops as distance increases. For both very large and very small $\S$ values, the forces are either attractive or repulsive regardless of the distance. However, for intermediate $S$ values (e.g. $\S=5$; yellow curves), there is a sign change in both the axial and transverse configurations as we move along the distance axis, indicating a reversal in the corresponding force. The location of the force reversal is an equilibrium point. The equilibrium is unstable along the axial direction, while it is stable along the transverse axis. For moderately large values of $\S$ (between around 5 and 15) the equilibrium point moves towards smaller $D$, where the two-term multipole theory is less accurate. 

The force also shows significant variation with $\S$ at fixed distance. For example, the axial configuration in figure \ref{FabreCompare} shows that the force starts  attractive for small $\S$, but flips sign and becomes repulsive at large $\S$. For large $\S$, all the curves appear to approach an ``envelope'' curve, deviating from  it only at small $D$. 

Figure \ref{FabreCompare}(c) shows the ``reorienting'' component of the force, which is positive for all $\S$ small to moderate $D$. At larger $D$, $F_{AT}$ decays rapidly while also changing sign, though this reversal is rather subtle. From \eqref{eqn.Fabre} we see that positive $F_{AT}$ corresponds to forces tending to reorient the particles into an orientation that is transverse to the oscillation axis. 

\subsection{Analytic approximations for $D \gg  \delta$}
The advantage of the formulation \eqref{RT2} is that it allows us to gain analytic insight into the behavior of the force. We focus on the limit of large separation between the particles ($D \gg 1$) and further consider non-overlapping Stokes layers ($D \gg  \delta$, or $D^2 \S \gg 1$). Recalling \eqref{RT2}, the force integral needs three inputs,
\begin{subequations}\label{eqn.3comps}
    \begin{align}
        \bv&=\bV^\infty+\bv_1^d+\bv_2^d,\\           
        \bm{\omega}&=\bm{\omega}_1^d+\bm{\omega}_2^d, \qmbox{and}\\
        \hat{\bv}&=\hat{\bv}_1^d+\hat{\bv}_2^d.
    \end{align}
\end{subequations}  
The oscillatory vorticity decays exponentially outside Stokes layers of thickness $\delta$ around each particle. Consequently, the integral in \eqref{RT2} splits into two ``localized'' integrals around each particle. We take advantage of the symmetry to evaluate the integral in the volume surrounding just of the particles (say particle $1$);  the integral in the volume surrounding particle 2 is identical. To perform the integral centered at particle 1, we approximate the primary disturbance and auxiliary velocity fields created by particle $2$, $\bv_2^d$ and $\hat{\bv}_2^d$, by a Taylor expansion around the center of particle $1$, retaining terms up to $O(D^{-4})$. Since the vorticity decays exponentially, we do not need to involve $\bm{\omega}_2$ in the integral around particle 1 in the regime where Stokes layers do not overlap. 

With these approximations, we perform the integral in spherical coordinates around particle 1 using \emph{Mathematica}. This yields analytic expressions for the three $F$ components in \eqref{eqn.Fabre}, valid in the limit $D \gg 1$, $D \gg \S^{-1/2}$. These force expressions take the general shape 
\begin{align}\label{eqn.forcegeneral}
    F_i=\frac{\A_{2i}}{D^2}+\frac{\A_{3i}}{D^3}+\frac{\A_{4i}}{D^4} + O(D^{-5}),
    \end{align}
where $i$ denotes $AA$, $TT$, or  $AT$, and the $\A_{ni}$ coefficients (9 in total) depend only on $\S$.
\begin{figure}
    \centering
    \includegraphics[width=0.9\textwidth]{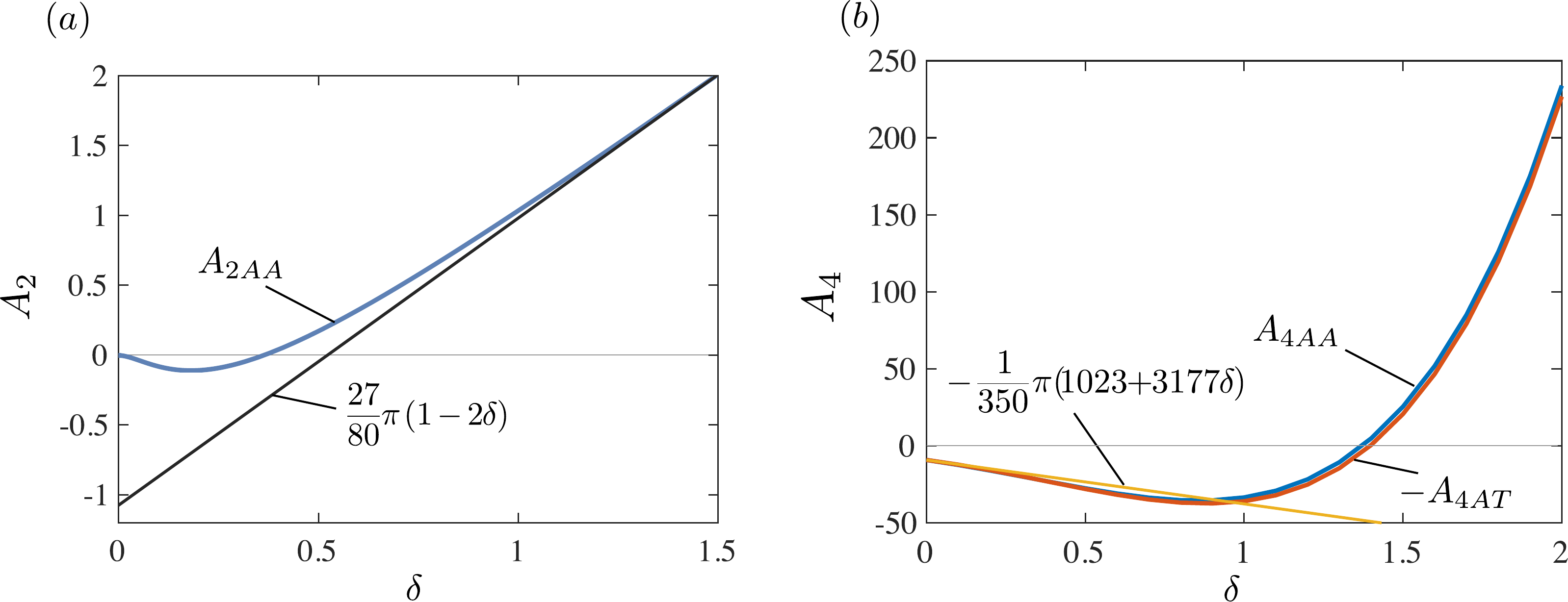}
    \caption{Analytic force coefficients $A_{2}$ and $A_{4}$ with respect to $\delta=\sqrt{2/\S}$. (a) Real part of $A_{2AA}$ and its asymptote for large $\delta$. Note that at approximately $\delta=0.35$, the value of $A_{2AA}$ is zero. (b) $A_{4AA}$ and $-A_{4AT}$ (both are nearly identical) and their asymptote for small $\delta$.}
    \label{fig.Afuncs}
\end{figure} 
While we do not provide detailed expressions for the $\A$ coefficients we identify important relationships between them. First we find that $\A_{3i} = \frac{3}{2}\A_{2i}$ for all configurations. We also find that all $AA$ and $TT$ coefficients have similar shapes but opposite signs, specifically $\A_{iTT} = -\frac{1}{2} \A_{iAA}$ for $i=\{1, 2, 3\}$. Finally, we find that $\A_{2AT} = \A_{3AT} = 0$, and that $\A_{4AT} \approx -\A_{4AA}$. Thus, \eqref{eqn.forcegeneral} simplifies as 
\begin{subequations}\label{eqn.forceapprox}
    \begin{align}
    F_{AA}&=\frac{\A_{2}}{D^2}\left(1+\frac{3}{2D}\right)+\frac{\A_{4}}{D^4}\\
    F_{TT}&=-\frac{\A_{2}}{2 D^2}\left(1+\frac{3}{2D}\right)-\frac{\A_{4}}{2D^4}\\
    F_{AT}&= \frac{\A_{4AT}}{D^4} \approx -\frac{\A_{4}}{D^4},
    \end{align}
\end{subequations}
where we have used the shorthand notation $A_2 = A_{2AA}$ and $A_4 = A_{4 AA}$. The simplified formulation \eqref{eqn.forceapprox} involves only two distinct coefficients, plotted in figure \ref{fig.Afuncs}. We see that all $\A$ coefficients reverse sign with $\S$, and that the two $\A_{4}$ coefficients are, for practical purposes, identical up to sign.

The leading contribution scales as $D^{-2}$. We interpret is as the drag force associated with a particle held fixed in the streaming flow created by oscillations around the other particle \cite{li2023structure}. The $(1 + 3/2D)$ correction factor is explained quantitatively by hydrodynamic interactions between approaching particles in viscous flows; see \cite{bre61_plane, ral17_hydroforce}. The $D^{-4}$ terms are more complicated, and arise from a combination of (i) Fax\'{e}n corrections and hydrodynamics interactions (viscous) due to particles suspended in the streaming field, and (ii) an inertio-viscous generalization of secondary radiation forces due to a product of the oscillatory flow velocity (dominated by the ambient oscillation) and its gradient ($\propto D^{-4}$) felt by each particle \cite{zhang2024particle}.

\begin{figure}
    \centering
    \includegraphics[width=1.00\textwidth]{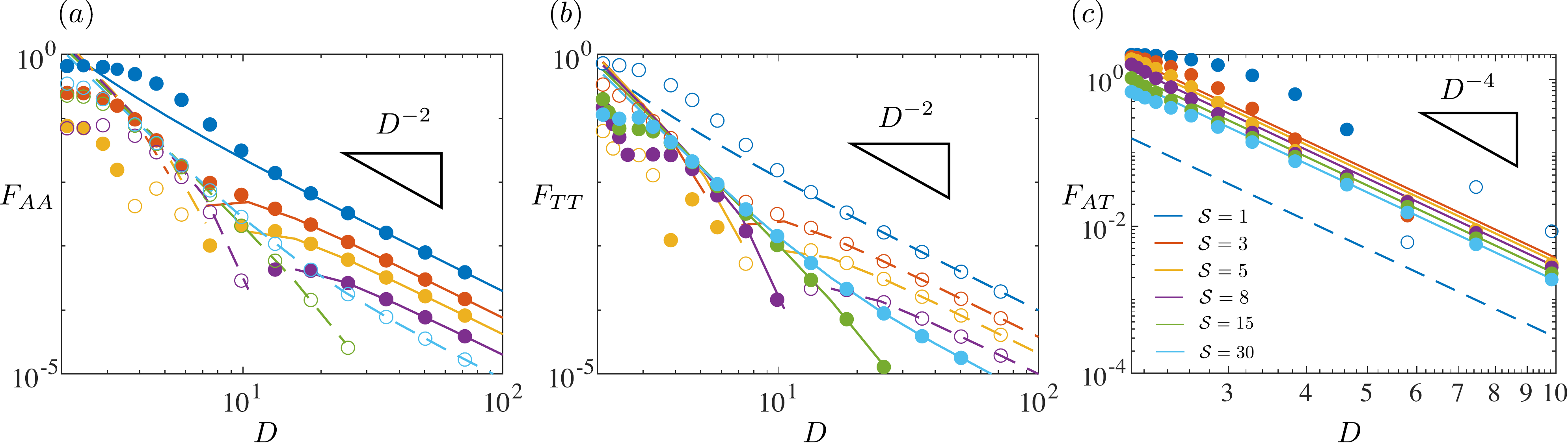}
    \caption{Comparison between the results of the semi-analytic calculations (symbols) and the analytic expression \eqref{eqn.forceapprox} (curves). Solid lines and filled circles represent positive values, while dashed lines and open circles represent negative values. }
    \label{LargeDCompare}
\end{figure}

Figure \ref{LargeDCompare} shows the analytical large-$D$ results obtained from \eqref{eqn.forceapprox} (curves) against the semi-analytic calculations (symbols). Positive values are indicated as solid curves and filled symbols, while negative values are indicated as dashed curves and open symbols. We see that the analytic theory reproduces the calculations for large $D$ with quantitative precision. Forces along the axis connecting the particles (attraction/repulsion) decay as $D^{-2}$, while the force normal to the axis ($F_{AT}$) decays more quickly as $D^{-4}$. 

From \eqref{fig.Afuncs}, we can also get a sense of how the force changes its sign. At large $D$, the $\A_2$ terms dominate, so the reversal of sign of the force is associated with the change in sign of $\A_{2}$. From figure \ref{fig.Afuncs} we see that this sign reversal occurs around $\delta=0.35$ ($\S \approx 16.3$). For large $\delta$ (small $\S$), we find the asymptotic behavior $A_{2} \sim \frac{27}{80}\pi\left(1-2\delta\right)$, whereas in the inviscid limit $\delta \to  0$ ($\S \to  \infty$), $\A_{2}$ approaches zero. By contrast, the $\A_{4}$ coefficient asymptotes to $-\frac{1}{350}\pi\left(1023+3177\delta\right)$ at small $\delta$. Notably, $\A_{4}$ does not vanish as $\delta \to 0$ and thus becomes the leading term in this limit. We also see that $\A_{4}$ is about two orders of magnitude greater than $\A_2$ for $O(1)$ values $\delta$, suggesting non--trivial convergence of the large-$D$ expansions \eqref{eqn.forceapprox}. 

\begin{figure}
    \centering
    \includegraphics[width=1.00\textwidth]{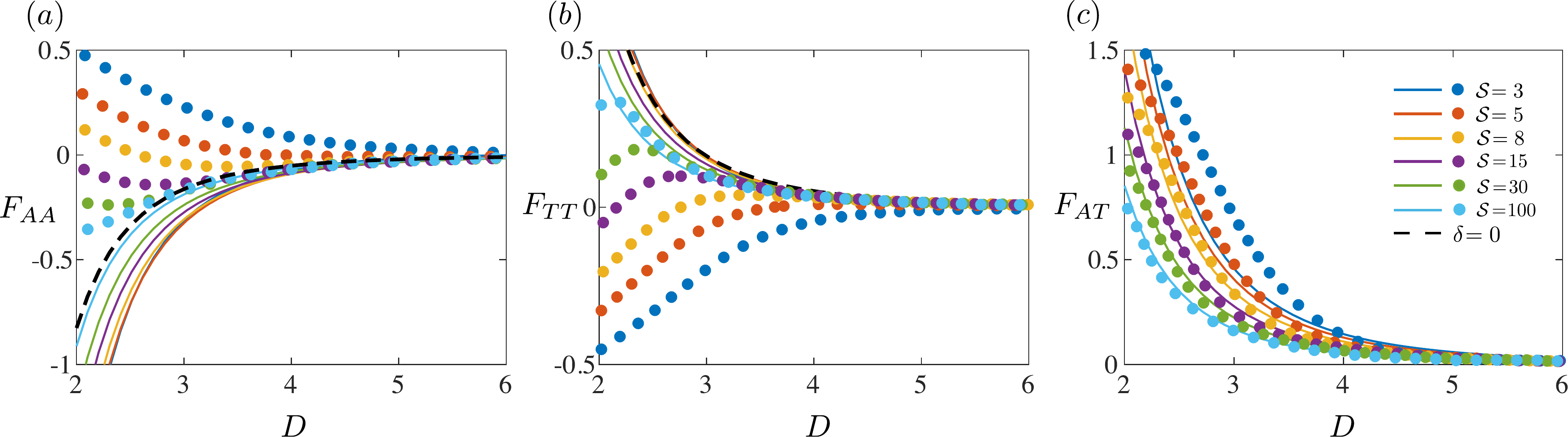}
    \caption{Comparison between the numerical calculations of \cite{fabre2017acoustic} (circles) and the analytic result \eqref{eqn.forceapprox} (solid curves). The dashed curve is the limit $\delta=0$, where the $D^{-4}$ term is dominant.}
    \label{SmallDCompare}
\end{figure}

For smaller distances $D = O(1)$, the analytic theory is strained as the Taylor and multipole expansions are less accurate, and because the Stokes layers are more likely to overlap. It nonetheless produces useful insights. Comparisons between the analytic theory (curves) and the computations of \cite{fabre2017acoustic} (symbols) are shown for $D$ between 2 and 6 in figure \ref{SmallDCompare}. Since the theory is only valid for $D \gg \delta$, it becomes restricted to $\delta \ll 1$ (or $\S \gg 1$) when separation distances become comparable to the particle radius ($D = O(1)$). The theoretical $F_{AA}$ and $F_{TT}$ coefficients corroborate this expectation, becoming more accurate at large $\S$ (figure \ref{SmallDCompare}). In the inviscid limit $\S \to  \infty$, the $A_2$ contribution vanishes while the $A_4$ contribution survives. The dashed curves in figure \ref{SmallDCompare}(a, b) represents this limit of the theory, which appears to approximately define an envelope of the data. The data ``peel off'' from this envelope at different $D$ (at smaller $D$ for larger $\S$), which we expect is the result of overlapping Stokes layers. It is interesting to note that the analytic theory captures that transverse force contribution $F_{AT}$ for large $\S$ all the way down to contact, while it is somewhat less accurate for smaller $\S$. 

\section{Freely suspended particles} \label{SecFreelySuspended}
We have so far assumed that the particles are fixed, and calculated the time-averaged inertial forces. We now relax this assumption and consider the case where the particles are freely suspended. They are now free to oscillate in response to the ambient flow, and also acquire time-averaged drift velocities from to a balance of time-averaged forces with viscous drag.

We consider particles of identical size and density $\rho_p$. Due to symmetry they oscillate at identical velocity, in phase with each other, when subjected to oscillatory ambient flow; see also \cite{Kleischmann_Luzzatto-Fegiz_Meiburg_Vowinckel_2024}. We denote this particle velocity (in dimensional terms) by $\bm{V}_{p1} \propto e^{i \omega t}$, and place the frame of reference at the instantaneous center of one of the particles (say particle 1) when solving for disturbance flow. In the particle-attached reference frame, the ambient flow becomes effectively replaced by the relative velocity $\bm{V}^{\infty} - \bm{V}_{p1}$.  To calculate this relative velocity between the particle and the flow, we invoke conservation of momentum of the oscillating particle. For this purpose we approximate each particle as being isolated, neglecting the $O(D^{-3})$ corrections to the oscillating velocity due to interactions between the particles. This approximation only leads to $O(D^{-5})$ corrections in the time-averaged motion, which is beyond the accuracy of the 2-term multipole expansion used here. Then, we invoke known results \cite{settnes2012forces, zhang2024particle} to find 
\begin{align}
    \bV_{p1} - \bV^{\infty} = \mathcal{R}\bV^{\infty}, \qmbox{with} \mathcal{R}=\frac{2\lambda^2\left(\tilde{\rho}-1\right)}{\lambda^2\left(2\tilde{\rho}+1\right)+9\lambda+9},
\end{align}
where $\tilde{\rho} = \rho_p/\rho$ is the density ratio between the particles and the fluid. The oscillating flow (both ambient and disturbance) in the co-moving reference frame is therefore identical to the one obtained for fixed particles, but for the mapping $\bV^{\infty} \mapsto  -\mathcal{R}\bV^{\infty}$. 
 
\begin{figure}
    \centering
    \includegraphics[width=1.00\textwidth]{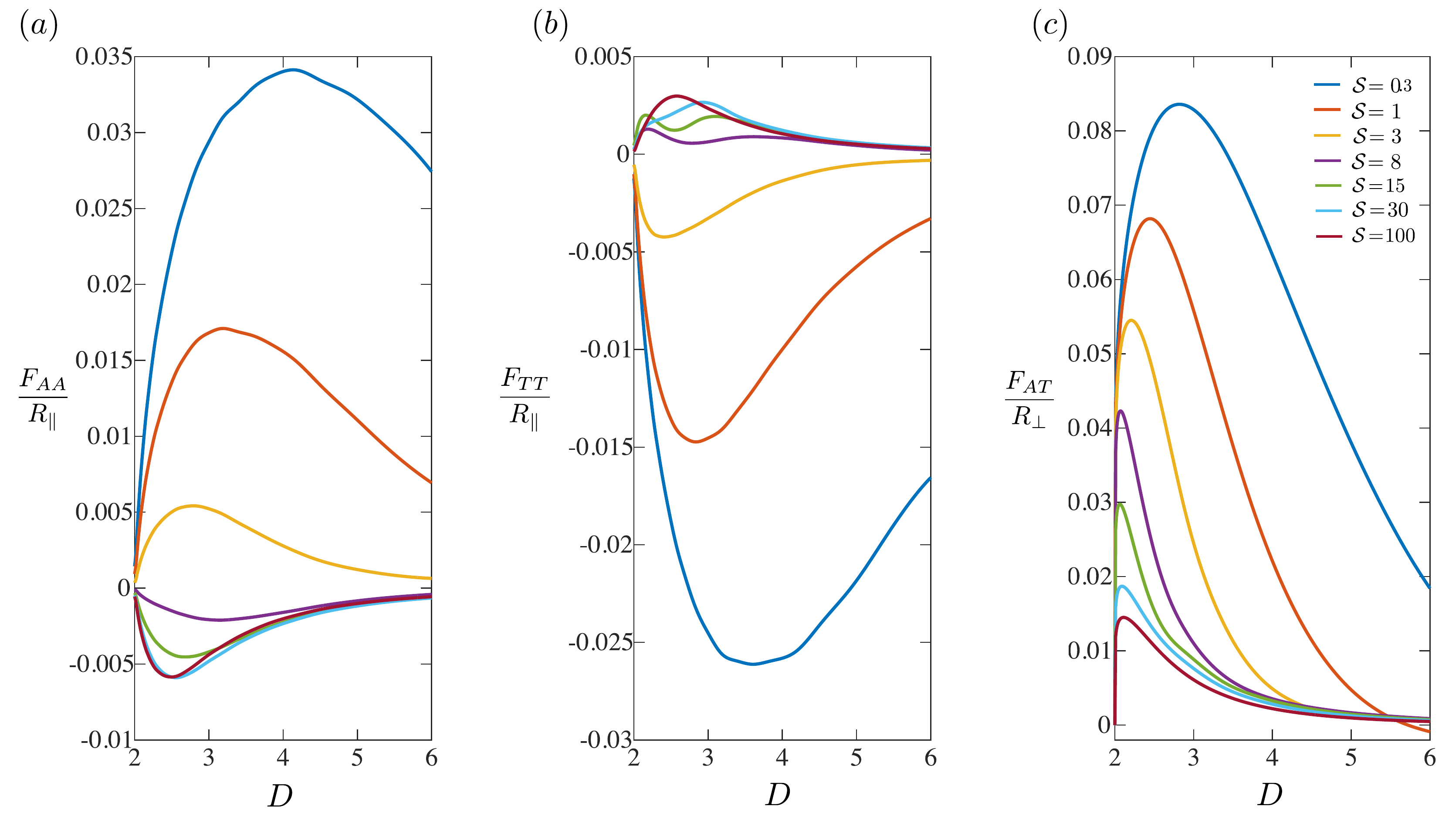}
    \caption{Secondary velocities corresponding to different configurations: (a) axial velocity due to axial oscillations ($AA$), (b) axial velocities due to transverse oscillations ($TT$), (c) transverse velocities ($AT$) corresponding to a reorientation of the particles with respect to the flow.}
    \label{FigV2}
\end{figure}
In addition to this change in the oscillatory flow, the particles are also able to execute a drift over many cycles, i.e.  $\left<\bV_2\right>$ is now nonzero. If the particles are allowed to move without the application of an external force ($\avg{\bm{F}} = 0$), the left-hand-side of \eqref{RT2} becomes $- \hat{\bm{F}}\cdot \avg{\bm{V}_2}$, while the right-hand-side is unchanged from before. The force in the auxiliary problem $\hat{\bm{F}}$ is related to the auxiliary velocity $\hat{\bm{V}}$ according to $\hat{\bm{F}}=-\mathbb{R}\cdot\hat{\bm{V}}$, where $\mathbb{R}=\mu a \left(R_{\parallel}\bm{e_{\parallel}}\bm{e_{\parallel}}+R_{\perp}\bm{e_{\perp}}\bm{e_{\perp}}\right)$ is a Stokes resistance matrix. The resistance coefficients $R_{\parallel}$ and $R_{\perp}$ correspond to steady motion of particles along and perpendicular to their line-of-centers, and depend on the separation $D$ (figure \ref{fig.aux}). Although they can be computed within the 2-term multipole expansion by the Fax'{e}n relation $\hat{\bm{F}}=-6\pi\mu a\left(\hat{\bm{V}}-\hat{\bm{\mathcal{V}}}\right)$, such an expansion does not capture the singular force contributions from lubrication theory. Instead, we use exact and asymptotic results \cite{bre61_plane} to construct uniform approximations that recover both the large-separation and near-contact limits  (see Appendix). The right-hand side of \eqref{RT2} remains unchanged from before.

The general reciprocal expression given by \eqref{RT2}, for freely-suspended (force-free) particles, yields the secondary velocity of particle 1 as (in dimensional form) 
\begin{align}\label{v2}
    \avg{\bm{V}_2} = \frac{(V^{\infty})^2}{a \omega} \S |\mathcal{R}|^2 \left(\frac{F_{AA}}{R_{\parallel}}\cos^2{\theta}\bm{e_{\parallel}}+\frac{F_{TT}}{R_{\parallel}}\sin^2{\theta}\bm{e_{\parallel}}+\frac{F_{AT}}{R_{\perp}}\sin{\theta}\cos{\theta}\bm{e_{\perp}}\right).
\end{align} 
The velocity of particle $2$ is equal in magnitude and opposite in sign, and $F_{AA}$, $F_{TT}$ and $F_{AT}$ are force coefficients defined in \eqref{eqn.Fabre}. Thus, we obtain the secondary velocity with very little extra effort. The resistance coefficients $R_{\perp}$ and $R_{\parallel}$ depend on $D$ only, approaching $6 \pi$ for large $D$ and zero at contact. The factor of $\S$ in \eqref{v2} reflects the balance between the inertial force scale in \eqref{eqn.Fabre} and viscous drag.


In figure \ref{FigV2} we plot the time-averaged particle velocity with respect to distance under the same values of $\S$ chosen in the force plots. The magnitudes of the velocities are consistent with the corresponding forces, while the peak values are shifted further from the contact point $D = 2$. This is due to the fact that the resistance to motion becomes infinite at contact (where inertial forces remain finite), while the force coefficients decay with distance (with the resistance being finite). The location and stability of the equilibrium points of the motion are inherited from those of the forces.

To close this section,  we compare our results against \cite{Kleischmann_Luzzatto-Fegiz_Meiburg_Vowinckel_2024}, where the Navier--Stokes equations were solved numerically to simulate the coupled flow and particle motion over time. From the trajectory data in Figure 4 of \cite{Kleischmann_Luzzatto-Fegiz_Meiburg_Vowinckel_2024} (particles in the AA configuration), we extract the time-averaged particle velocities at 50 cycles from the initial condition (long enough to avoid transient effects). 

\begin{figure}
     \centering
     \includegraphics[width=0.5\textwidth]{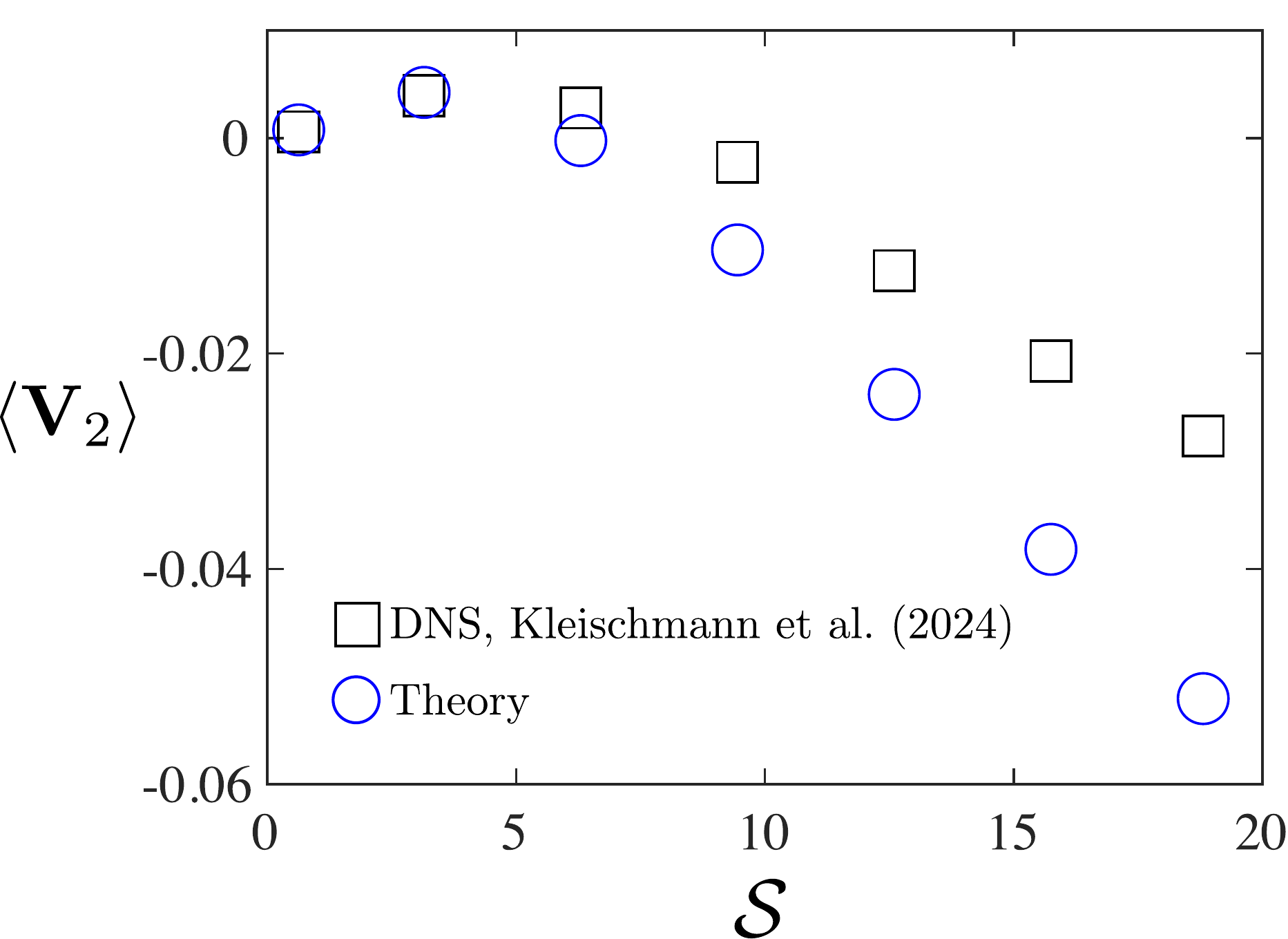}
     \caption{Comparison of secondary velocity on particle surface between the present theory (circles) and DNS trajectory data of \cite{Kleischmann_Luzzatto-Fegiz_Meiburg_Vowinckel_2024} (squares).}
     \label{kle}
\end{figure}

We analyze 7 sets of data ranging from $\S=0.63$ to $\S=18.81$ (the entire range considered in \cite{Kleischmann_Luzzatto-Fegiz_Meiburg_Vowinckel_2024}), which correspond to our small to moderate Stokes numbers, and plot the velocities in figure \ref{kle}, without additional fitting parameters. Due to the setup of these DNS, each velocity datum occurs at a slightly different $D$, which we match in the comparison to present theory. 
The theory is in good agreement with DNS. The direction of motion is faithfully captured by the theory for all $\S$. When $\S$ values are smaller, the velocity magnitudes are also captured accurately. The roughly linear increase of velocities (greater negative values in this case) at larger $\S$ is reproduced well by the theory, though the magnitude is overpredicted relative to the DNS. 
One possible cause for this is the proximity of the particles to the walls of the computational domain, which not only add additional resistance, but also affect the structure of the secondary ``streaming'' flow generated by the particles. The inertia of the secondary motion, which is neglected here, becomes larger with $\S$, and may lead to an increase in the drag experienced by the particles. Despite these differences, it is clear that the present theory provides a useful means to predict secondary particle motion with relatively little computational effort.   

\section{Conclusions} \label{SecConc}

To conclude, we have explored the inertio-viscous interactions between two identical particles in an oscillatory flow. Through a theoretical approach involving small amplitude expansions, multipole expansions, and the application of the Lorentz reciprocal theorem, we derived a method to calculate the time-averaged secondary forces acting between the particles. The analysis demonstrated that while the full nonlinear flow is complex, it is possible to reduce the problem to an analytically tractable form by focusing on vorticity effects near the particle surfaces. By using the reciprocal theorem, we circumvented the need for direct computation of secondary stress, leading to an efficient framework for understanding particle interactions in oscillatory flows. By providing new analytic insight, the present framework fills the gap between existing work on direct numerical solutions and numerical solutions of scale-separated equations from perturbation theory. Due to its low computational cost, the present work can cover a broader range of frequency and separation distances. 

The reversal of force indicates the existence of an equilibrium distance between particles exposed to oscillatory flow, which may be useful for explaining particle pattern formation when exposed to oscillatory fields. Although we focused on a single pair of particles for the theoretical development here, the framework has the potential to be extended to multiple particles of identical size and density. Another possibility for exploration is systems of particles with different sizes, shapes or densities.  Our findings not only enhance our understanding of particle dynamics in fluid systems but also provide useful theoretical tools for further exploration in practical applications.



\begin{acknowledgments}
The authors thank the National Science Foundation for support through award CBET-2143943.
\end{acknowledgments}

\appendix
\section{Resistance coefficients}
\begin{figure}
    \centering
    \includegraphics[width=1.00\textwidth]{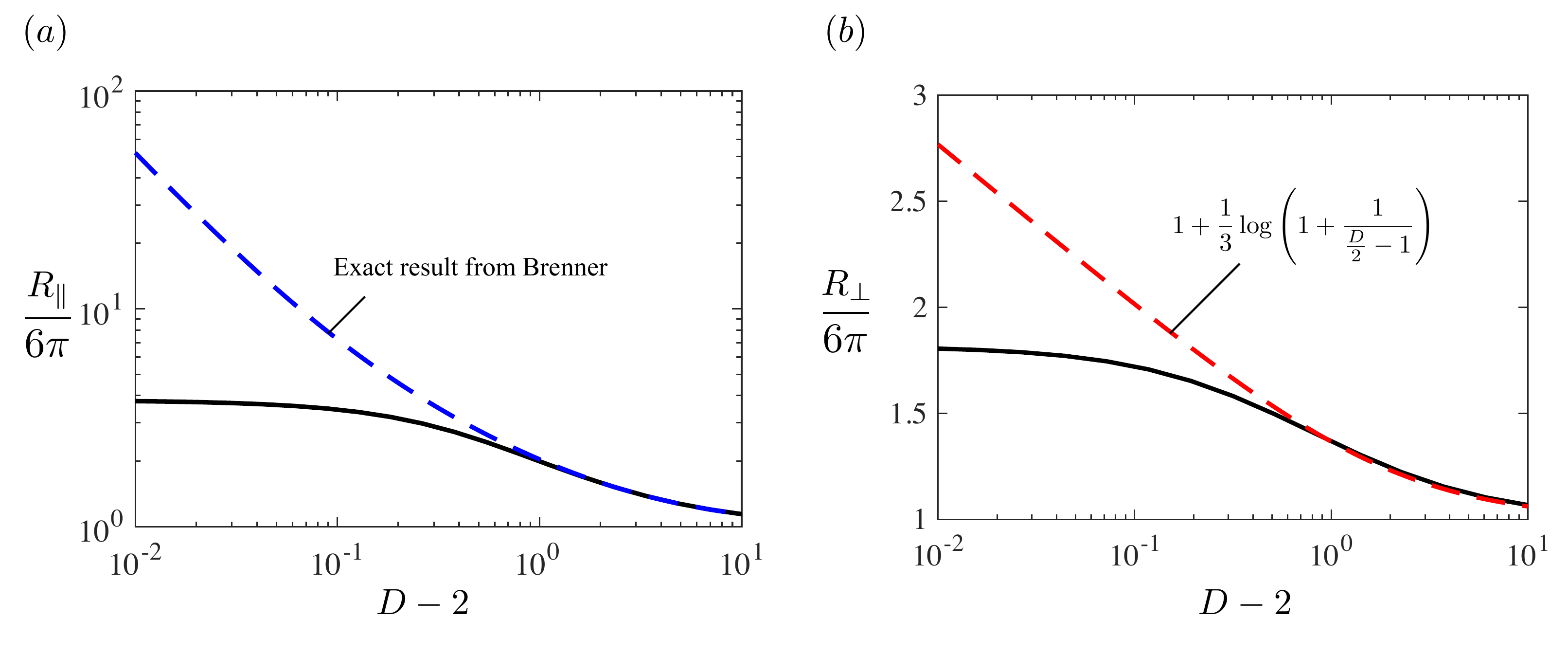}
    \caption{Stokes resistances showing the approximate large-$D$ result from the two-term multipole expansion with Fax\'{e}n's law (solid curves) and uniformly valid expansion (dashed curves). (a) Parallel configuration. The blue dashed curve denotes the exact result of \cite{bre61_plane}. (b)Perpendicular configuration. The red dashed curve denotes a universal approximation that accounts for lubrication theory. In both configuration we use the dashed curves, which are accurate across the entire range of inter-particle separation distances, to calculate the velocities in figure \ref{FigV2}.}
    \label{fig.resis.pdf}
\end{figure}

We plot the resistance coefficients for both axis-parallel and axis-perpendicular motion in figure \ref{fig.resis.pdf}. Calculations from Fax\'{e}n's law with the 2-term multipole expansion are indicated as solid curves. They are accurate at large distances $(D \gtrsim 3)$ but do not account for lubrication effects when particles are near contact. We remedy this by using a combination of known exact and asymptotic results. For $R_{\parallel}$ we use the exact result of  \cite{bre61_plane}, which becomes singular as the inverse of surface separation distance, $R_{\parallel} \sim 6 \pi (1/4) (D/2-1)^{-1}$, as the particles approach contact (figure \ref{fig.resis.pdf}(a)). The perpendicular resistance behaves as $R_{\perp} \sim 6 \pi (1/3) \log (D/2-1)^{-1}$ close to contact, and approaches $6 \pi$ for large $D$. We construct a uniform approximation that captures both of these limits, plotted as a dashed curve in figure \ref{fig.resis.pdf}(b). 


\begin{thebibliography}{33}%
    \makeatletter
    \providecommand \@ifxundefined [1]{%
     \@ifx{#1\undefined}
    }%
    \providecommand \@ifnum [1]{%
     \ifnum #1\expandafter \@firstoftwo
     \else \expandafter \@secondoftwo
     \fi
    }%
    \providecommand \@ifx [1]{%
     \ifx #1\expandafter \@firstoftwo
     \else \expandafter \@secondoftwo
     \fi
    }%
    \providecommand \natexlab [1]{#1}%
    \providecommand \enquote  [1]{``#1''}%
    \providecommand \bibnamefont  [1]{#1}%
    \providecommand \bibfnamefont [1]{#1}%
    \providecommand \citenamefont [1]{#1}%
    \providecommand \href@noop [0]{\@secondoftwo}%
    \providecommand \href [0]{\begingroup \@sanitize@url \@href}%
    \providecommand \@href[1]{\@@startlink{#1}\@@href}%
    \providecommand \@@href[1]{\endgroup#1\@@endlink}%
    \providecommand \@sanitize@url [0]{\catcode `\\12\catcode `\$12\catcode
      `\&12\catcode `\#12\catcode `\^12\catcode `\_12\catcode `\%12\relax}%
    \providecommand \@@startlink[1]{}%
    \providecommand \@@endlink[0]{}%
    \providecommand \url  [0]{\begingroup\@sanitize@url \@url }%
    \providecommand \@url [1]{\endgroup\@href {#1}{\urlprefix }}%
    \providecommand \urlprefix  [0]{URL }%
    \providecommand \Eprint [0]{\href }%
    \providecommand \doibase [0]{https://doi.org/}%
    \providecommand \selectlanguage [0]{\@gobble}%
    \providecommand \bibinfo  [0]{\@secondoftwo}%
    \providecommand \bibfield  [0]{\@secondoftwo}%
    \providecommand \translation [1]{[#1]}%
    \providecommand \BibitemOpen [0]{}%
    \providecommand \bibitemStop [0]{}%
    \providecommand \bibitemNoStop [0]{.\EOS\space}%
    \providecommand \EOS [0]{\spacefactor3000\relax}%
    \providecommand \BibitemShut  [1]{\csname bibitem#1\endcsname}%
    \let\auto@bib@innerbib\@empty
    \bibitem [{\citenamefont {Riley}(2001)}]{riley2001steady}%
      \BibitemOpen
      \bibfield  {author} {\bibinfo {author} {\bibfnamefont {N.}~\bibnamefont
      {Riley}},\ }\bibfield  {title} {\bibinfo {title} {Steady streaming},\
      }\href@noop {} {\bibfield  {journal} {\bibinfo  {journal} {Annu. Rev. Fluid.
      Mech.}\ }\textbf {\bibinfo {volume} {33}},\ \bibinfo {pages} {43} (\bibinfo
      {year} {2001})}\BibitemShut {NoStop}%
    \bibitem [{\citenamefont {King}(1934)}]{king1934acoustic}%
      \BibitemOpen
      \bibfield  {author} {\bibinfo {author} {\bibfnamefont {L.~V.}\ \bibnamefont
      {King}},\ }\bibfield  {title} {\bibinfo {title} {On the acoustic radiation
      pressure on spheres},\ }\href@noop {} {\bibfield  {journal} {\bibinfo
      {journal} {Proc. Roy. Soc. Lond. A}\ }\textbf {\bibinfo {volume} {147}},\
      \bibinfo {pages} {212} (\bibinfo {year} {1934})}\BibitemShut {NoStop}%
    \bibitem [{\citenamefont {{Gor'kov}}(1962)}]{Gorkov1962}%
      \BibitemOpen
      \bibfield  {author} {\bibinfo {author} {\bibfnamefont {L.~P.}\ \bibnamefont
      {{Gor'kov}}},\ }\bibfield  {title} {\bibinfo {title} {{On the Forces Acting
      on a Small Particle in an Acoustical Field in an Ideal Fluid}},\ }\href@noop
      {} {\bibfield  {journal} {\bibinfo  {journal} {Soviet Physics Doklady}\
      }\textbf {\bibinfo {volume} {6}},\ \bibinfo {pages} {773} (\bibinfo {year}
      {1962})}\BibitemShut {NoStop}%
    \bibitem [{\citenamefont {Settnes}\ and\ \citenamefont
      {Bruus}(2012)}]{settnes2012forces}%
      \BibitemOpen
      \bibfield  {author} {\bibinfo {author} {\bibfnamefont {M.}~\bibnamefont
      {Settnes}}\ and\ \bibinfo {author} {\bibfnamefont {H.}~\bibnamefont
      {Bruus}},\ }\bibfield  {title} {\bibinfo {title} {Forces acting on a small
      particle in an acoustical field in a viscous fluid},\ }\href@noop {}
      {\bibfield  {journal} {\bibinfo  {journal} {Phys. Rev. E}\ }\textbf {\bibinfo
      {volume} {85}},\ \bibinfo {pages} {016327} (\bibinfo {year}
      {2012})}\BibitemShut {NoStop}%
    \bibitem [{\citenamefont {Friend}\ and\ \citenamefont
      {Yeo}(2011)}]{friend2011microscale}%
      \BibitemOpen
      \bibfield  {author} {\bibinfo {author} {\bibfnamefont {J.}~\bibnamefont
      {Friend}}\ and\ \bibinfo {author} {\bibfnamefont {L.~Y.}\ \bibnamefont
      {Yeo}},\ }\bibfield  {title} {\bibinfo {title} {Microscale acoustofluidics:
      Microfluidics driven via acoustics and ultrasonics},\ }\href@noop {}
      {\bibfield  {journal} {\bibinfo  {journal} {Rev. Mod. Phys.}\ }\textbf
      {\bibinfo {volume} {83}},\ \bibinfo {pages} {647} (\bibinfo {year}
      {2011})}\BibitemShut {NoStop}%
    \bibitem [{\citenamefont {Mutlu}\ \emph {et~al.}(2018)\citenamefont {Mutlu},
      \citenamefont {Edd},\ and\ \citenamefont {Toner}}]{mutlu2018oscillatory}%
      \BibitemOpen
      \bibfield  {author} {\bibinfo {author} {\bibfnamefont {B.~R.}\ \bibnamefont
      {Mutlu}}, \bibinfo {author} {\bibfnamefont {J.~F.}\ \bibnamefont {Edd}},\
      and\ \bibinfo {author} {\bibfnamefont {M.}~\bibnamefont {Toner}},\ }\bibfield
       {title} {\bibinfo {title} {Oscillatory inertial focusing in infinite
      microchannels},\ }\href@noop {} {\bibfield  {journal} {\bibinfo  {journal}
      {Proc. Natl. Acad. Sci.}\ }\textbf {\bibinfo {volume} {115}},\ \bibinfo
      {pages} {7682} (\bibinfo {year} {2018})}\BibitemShut {NoStop}%
    \bibitem [{\citenamefont {Rufo}\ \emph {et~al.}(2022)\citenamefont {Rufo},
      \citenamefont {Cai}, \citenamefont {Friend}, \citenamefont {Wiklund},\ and\
      \citenamefont {Huang}}]{rufo2022acoustofluidics}%
      \BibitemOpen
      \bibfield  {author} {\bibinfo {author} {\bibfnamefont {J.}~\bibnamefont
      {Rufo}}, \bibinfo {author} {\bibfnamefont {F.}~\bibnamefont {Cai}}, \bibinfo
      {author} {\bibfnamefont {J.}~\bibnamefont {Friend}}, \bibinfo {author}
      {\bibfnamefont {M.}~\bibnamefont {Wiklund}},\ and\ \bibinfo {author}
      {\bibfnamefont {T.~J.}\ \bibnamefont {Huang}},\ }\bibfield  {title} {\bibinfo
      {title} {Acoustofluidics for biomedical applications},\ }\href@noop {}
      {\bibfield  {journal} {\bibinfo  {journal} {Nature Reviews Methods Primers}\
      }\textbf {\bibinfo {volume} {2}},\ \bibinfo {pages} {30} (\bibinfo {year}
      {2022})}\BibitemShut {NoStop}%
    \bibitem [{\citenamefont {Yang}\ \emph {et~al.}(2022)\citenamefont {Yang},
      \citenamefont {Tian}, \citenamefont {Wang}, \citenamefont {Rufo},
      \citenamefont {Li}, \citenamefont {Mai}, \citenamefont {Xia}, \citenamefont
      {Bachman}, \citenamefont {Huang}, \citenamefont {Wu} \emph
      {et~al.}}]{yang2022harmonic}%
      \BibitemOpen
      \bibfield  {author} {\bibinfo {author} {\bibfnamefont {S.}~\bibnamefont
      {Yang}}, \bibinfo {author} {\bibfnamefont {Z.}~\bibnamefont {Tian}}, \bibinfo
      {author} {\bibfnamefont {Z.}~\bibnamefont {Wang}}, \bibinfo {author}
      {\bibfnamefont {J.}~\bibnamefont {Rufo}}, \bibinfo {author} {\bibfnamefont
      {P.}~\bibnamefont {Li}}, \bibinfo {author} {\bibfnamefont {J.}~\bibnamefont
      {Mai}}, \bibinfo {author} {\bibfnamefont {J.}~\bibnamefont {Xia}}, \bibinfo
      {author} {\bibfnamefont {H.}~\bibnamefont {Bachman}}, \bibinfo {author}
      {\bibfnamefont {P.-H.}\ \bibnamefont {Huang}}, \bibinfo {author}
      {\bibfnamefont {M.}~\bibnamefont {Wu}}, \emph {et~al.},\ }\bibfield  {title}
      {\bibinfo {title} {Harmonic acoustics for dynamic and selective particle
      manipulation},\ }\href@noop {} {\bibfield  {journal} {\bibinfo  {journal}
      {Nature Mat.}\ }\textbf {\bibinfo {volume} {21}},\ \bibinfo {pages} {540}
      (\bibinfo {year} {2022})}\BibitemShut {NoStop}%
    \bibitem [{\citenamefont {Foresti}\ and\ \citenamefont
      {Poulikakos}(2014)}]{foresti2014acoustophoretic}%
      \BibitemOpen
      \bibfield  {author} {\bibinfo {author} {\bibfnamefont {D.}~\bibnamefont
      {Foresti}}\ and\ \bibinfo {author} {\bibfnamefont {D.}~\bibnamefont
      {Poulikakos}},\ }\bibfield  {title} {\bibinfo {title} {Acoustophoretic
      contactless elevation, orbital transport and spinning of matter in air},\
      }\href@noop {} {\bibfield  {journal} {\bibinfo  {journal} {Phys. Rev. Lett.}\
      }\textbf {\bibinfo {volume} {112}},\ \bibinfo {pages} {024301} (\bibinfo
      {year} {2014})}\BibitemShut {NoStop}%
    \bibitem [{\citenamefont {Andrade}\ \emph {et~al.}(2020)\citenamefont
      {Andrade}, \citenamefont {Marzo},\ and\ \citenamefont
      {Adamowski}}]{andrade2020acoustic}%
      \BibitemOpen
      \bibfield  {author} {\bibinfo {author} {\bibfnamefont {M.~A.}\ \bibnamefont
      {Andrade}}, \bibinfo {author} {\bibfnamefont {A.}~\bibnamefont {Marzo}},\
      and\ \bibinfo {author} {\bibfnamefont {J.~C.}\ \bibnamefont {Adamowski}},\
      }\bibfield  {title} {\bibinfo {title} {Acoustic levitation in mid-air: Recent
      advances, challenges, and future perspectives},\ }\href@noop {} {\bibfield
      {journal} {\bibinfo  {journal} {Appl. Phys. Lett.}\ }\textbf {\bibinfo
      {volume} {116}},\ \bibinfo {pages} {250501} (\bibinfo {year}
      {2020})}\BibitemShut {NoStop}%
    \bibitem [{\citenamefont {Lee}\ \emph {et~al.}(2018)\citenamefont {Lee},
      \citenamefont {James}, \citenamefont {Waitukaitis},\ and\ \citenamefont
      {Jaeger}}]{lee2018collisional}%
      \BibitemOpen
      \bibfield  {author} {\bibinfo {author} {\bibfnamefont {V.}~\bibnamefont
      {Lee}}, \bibinfo {author} {\bibfnamefont {N.~M.}\ \bibnamefont {James}},
      \bibinfo {author} {\bibfnamefont {S.~R.}\ \bibnamefont {Waitukaitis}},\ and\
      \bibinfo {author} {\bibfnamefont {H.~M.}\ \bibnamefont {Jaeger}},\ }\bibfield
       {title} {\bibinfo {title} {Collisional charging of individual submillimeter
      particles: Using ultrasonic levitation to initiate and track charge
      transfer},\ }\href@noop {} {\bibfield  {journal} {\bibinfo  {journal} {Phys.
      Rev. Mat.}\ }\textbf {\bibinfo {volume} {2}},\ \bibinfo {pages} {035602}
      (\bibinfo {year} {2018})}\BibitemShut {NoStop}%
    \bibitem [{\citenamefont {Lirette}\ \emph {et~al.}(2019)\citenamefont
      {Lirette}, \citenamefont {Mobley},\ and\ \citenamefont
      {Zhang}}]{lirette2019ultrasonic}%
      \BibitemOpen
      \bibfield  {author} {\bibinfo {author} {\bibfnamefont {R.}~\bibnamefont
      {Lirette}}, \bibinfo {author} {\bibfnamefont {J.}~\bibnamefont {Mobley}},\
      and\ \bibinfo {author} {\bibfnamefont {L.}~\bibnamefont {Zhang}},\ }\bibfield
       {title} {\bibinfo {title} {Ultrasonic extraction and manipulation of
      droplets from a liquid-liquid interface with near-field acoustic tweezers},\
      }\href@noop {} {\bibfield  {journal} {\bibinfo  {journal} {Phys. Rev. Appl.}\
      }\textbf {\bibinfo {volume} {12}},\ \bibinfo {pages} {061001} (\bibinfo
      {year} {2019})}\BibitemShut {NoStop}%
    \bibitem [{\citenamefont {Klotsa}\ \emph {et~al.}(2009)\citenamefont {Klotsa},
      \citenamefont {Swift}, \citenamefont {Bowley},\ and\ \citenamefont
      {King}}]{Klotsa_2009}%
      \BibitemOpen
      \bibfield  {author} {\bibinfo {author} {\bibfnamefont {D.}~\bibnamefont
      {Klotsa}}, \bibinfo {author} {\bibfnamefont {M.~R.}\ \bibnamefont {Swift}},
      \bibinfo {author} {\bibfnamefont {R.~M.}\ \bibnamefont {Bowley}},\ and\
      \bibinfo {author} {\bibfnamefont {P.~J.}\ \bibnamefont {King}},\ }\bibfield
      {title} {\bibinfo {title} {Chain formation of spheres in oscillatory fluid
      flows},\ }\href {https://doi.org/10.1103%2Fphysreve.79.021302} {\bibfield
      {journal} {\bibinfo  {journal} {Phys. Rev. E}\ }\textbf {\bibinfo {volume}
      {79}} (\bibinfo {year} {2009})}\BibitemShut {NoStop}%
    \bibitem [{\citenamefont {Voth}\ \emph {et~al.}(2002)\citenamefont {Voth},
      \citenamefont {Bigger}, \citenamefont {Buckley}, \citenamefont {Losert},
      \citenamefont {Brenner}, \citenamefont {Stone},\ and\ \citenamefont
      {Gollub}}]{voth2002ordered}%
      \BibitemOpen
      \bibfield  {author} {\bibinfo {author} {\bibfnamefont {G.~A.}\ \bibnamefont
      {Voth}}, \bibinfo {author} {\bibfnamefont {B.}~\bibnamefont {Bigger}},
      \bibinfo {author} {\bibfnamefont {M.}~\bibnamefont {Buckley}}, \bibinfo
      {author} {\bibfnamefont {W.}~\bibnamefont {Losert}}, \bibinfo {author}
      {\bibfnamefont {M.}~\bibnamefont {Brenner}}, \bibinfo {author} {\bibfnamefont
      {H.~A.}\ \bibnamefont {Stone}},\ and\ \bibinfo {author} {\bibfnamefont
      {J.}~\bibnamefont {Gollub}},\ }\bibfield  {title} {\bibinfo {title} {Ordered
      clusters and dynamical states of particles in a vibrated fluid},\ }\href@noop
      {} {\bibfield  {journal} {\bibinfo  {journal} {Phys. Rev. Lett.}\ }\textbf
      {\bibinfo {volume} {88}},\ \bibinfo {pages} {234301} (\bibinfo {year}
      {2002})}\BibitemShut {NoStop}%
    \bibitem [{\citenamefont {Sazhin}\ \emph {et~al.}(2008)\citenamefont {Sazhin},
      \citenamefont {Shakked}, \citenamefont {Sobolev},\ and\ \citenamefont
      {Katoshevski}}]{sazhin2008particle}%
      \BibitemOpen
      \bibfield  {author} {\bibinfo {author} {\bibfnamefont {S.}~\bibnamefont
      {Sazhin}}, \bibinfo {author} {\bibfnamefont {T.}~\bibnamefont {Shakked}},
      \bibinfo {author} {\bibfnamefont {V.}~\bibnamefont {Sobolev}},\ and\ \bibinfo
      {author} {\bibfnamefont {D.}~\bibnamefont {Katoshevski}},\ }\bibfield
      {title} {\bibinfo {title} {Particle grouping in oscillating flows},\
      }\href@noop {} {\bibfield  {journal} {\bibinfo  {journal} {European Journal
      of Mechanics-B/Fluids}\ }\textbf {\bibinfo {volume} {27}},\ \bibinfo {pages}
      {131} (\bibinfo {year} {2008})}\BibitemShut {NoStop}%
    \bibitem [{\citenamefont {Lim}\ \emph {et~al.}(2019)\citenamefont {Lim},
      \citenamefont {Souslov}, \citenamefont {Vitelli},\ and\ \citenamefont
      {Jaeger}}]{lim2019cluster}%
      \BibitemOpen
      \bibfield  {author} {\bibinfo {author} {\bibfnamefont {M.~X.}\ \bibnamefont
      {Lim}}, \bibinfo {author} {\bibfnamefont {A.}~\bibnamefont {Souslov}},
      \bibinfo {author} {\bibfnamefont {V.}~\bibnamefont {Vitelli}},\ and\ \bibinfo
      {author} {\bibfnamefont {H.~M.}\ \bibnamefont {Jaeger}},\ }\bibfield  {title}
      {\bibinfo {title} {Cluster formation by acoustic forces and active
      fluctuations in levitated granular matter},\ }\href@noop {} {\bibfield
      {journal} {\bibinfo  {journal} {Nature Phys.}\ }\textbf {\bibinfo {volume}
      {15}},\ \bibinfo {pages} {460} (\bibinfo {year} {2019})}\BibitemShut
      {NoStop}%
    \bibitem [{\citenamefont {Ingber}\ and\ \citenamefont
      {Vorobieff}(2013)}]{ingber2013particle}%
      \BibitemOpen
      \bibfield  {author} {\bibinfo {author} {\bibfnamefont {M.}~\bibnamefont
      {Ingber}}\ and\ \bibinfo {author} {\bibfnamefont {P.}~\bibnamefont
      {Vorobieff}},\ }\bibfield  {title} {\bibinfo {title} {Particle interactions
      in oscillatory stokes flow},\ }\href@noop {} {\bibfield  {journal} {\bibinfo
      {journal} {WIT Transactions on Engineering Sciences}\ }\textbf {\bibinfo
      {volume} {79}},\ \bibinfo {pages} {147} (\bibinfo {year} {2013})}\BibitemShut
      {NoStop}%
    \bibitem [{\citenamefont {Fabre}\ \emph {et~al.}(2017)\citenamefont {Fabre},
      \citenamefont {Jalal}, \citenamefont {Leontini},\ and\ \citenamefont
      {Manasseh}}]{fabre2017acoustic}%
      \BibitemOpen
      \bibfield  {author} {\bibinfo {author} {\bibfnamefont {D.}~\bibnamefont
      {Fabre}}, \bibinfo {author} {\bibfnamefont {J.}~\bibnamefont {Jalal}},
      \bibinfo {author} {\bibfnamefont {J.~S.}\ \bibnamefont {Leontini}},\ and\
      \bibinfo {author} {\bibfnamefont {R.}~\bibnamefont {Manasseh}},\ }\bibfield
      {title} {\bibinfo {title} {Acoustic streaming and the induced forces between
      two spheres},\ }\href@noop {} {\bibfield  {journal} {\bibinfo  {journal}
      {Journal of Fluid Mechanics}\ }\textbf {\bibinfo {volume} {810}},\ \bibinfo
      {pages} {378} (\bibinfo {year} {2017})}\BibitemShut {NoStop}%
    \bibitem [{\citenamefont {Klotsa}\ \emph {et~al.}(2007)\citenamefont {Klotsa},
      \citenamefont {Swift}, \citenamefont {Bowley},\ and\ \citenamefont
      {King}}]{klotsa2007interaction}%
      \BibitemOpen
      \bibfield  {author} {\bibinfo {author} {\bibfnamefont {D.}~\bibnamefont
      {Klotsa}}, \bibinfo {author} {\bibfnamefont {M.~R.}\ \bibnamefont {Swift}},
      \bibinfo {author} {\bibfnamefont {R.}~\bibnamefont {Bowley}},\ and\ \bibinfo
      {author} {\bibfnamefont {P.}~\bibnamefont {King}},\ }\bibfield  {title}
      {\bibinfo {title} {Interaction of spheres in oscillatory fluid flows},\
      }\href@noop {} {\bibfield  {journal} {\bibinfo  {journal} {Physical Review
      E}\ }\textbf {\bibinfo {volume} {76}},\ \bibinfo {pages} {056314} (\bibinfo
      {year} {2007})}\BibitemShut {NoStop}%
    \bibitem [{\citenamefont {Kleischmann}\ \emph {et~al.}(2024)\citenamefont
      {Kleischmann}, \citenamefont {Luzzatto-Fegiz}, \citenamefont {Meiburg},\ and\
      \citenamefont
      {Vowinckel}}]{Kleischmann_Luzzatto-Fegiz_Meiburg_Vowinckel_2024}%
      \BibitemOpen
      \bibfield  {author} {\bibinfo {author} {\bibfnamefont {F.}~\bibnamefont
      {Kleischmann}}, \bibinfo {author} {\bibfnamefont {P.}~\bibnamefont
      {Luzzatto-Fegiz}}, \bibinfo {author} {\bibfnamefont {E.}~\bibnamefont
      {Meiburg}},\ and\ \bibinfo {author} {\bibfnamefont {B.}~\bibnamefont
      {Vowinckel}},\ }\bibfield  {title} {\bibinfo {title} {Pairwise interaction of
      spherical particles aligned in high-frequency oscillatory flow},\ }\href
      {https://doi.org/10.1017/jfm.2024.251} {\bibfield  {journal} {\bibinfo
      {journal} {Journal of Fluid Mechanics}\ }\textbf {\bibinfo {volume} {984}},\
      \bibinfo {pages} {A57} (\bibinfo {year} {2024})}\BibitemShut {NoStop}%
    \bibitem [{\citenamefont {Zhang}\ and\ \citenamefont
      {Marston}(2014)}]{zhang2014acoustic}%
      \BibitemOpen
      \bibfield  {author} {\bibinfo {author} {\bibfnamefont {L.}~\bibnamefont
      {Zhang}}\ and\ \bibinfo {author} {\bibfnamefont {P.~L.}\ \bibnamefont
      {Marston}},\ }\bibfield  {title} {\bibinfo {title} {Acoustic radiation torque
      on small objects in viscous fluids and connection with viscous dissipation},\
      }\href@noop {} {\bibfield  {journal} {\bibinfo  {journal} {The Journal of the
      Acoustical Society of America}\ }\textbf {\bibinfo {volume} {136}},\ \bibinfo
      {pages} {2917} (\bibinfo {year} {2014})}\BibitemShut {NoStop}%
    \bibitem [{\citenamefont {Agarwal}\ \emph {et~al.}(2021)\citenamefont
      {Agarwal}, \citenamefont {Chan}, \citenamefont {Rallabandi}, \citenamefont
      {Gazzola},\ and\ \citenamefont {Hilgenfeldt}}]{agarwal2021unrecognized}%
      \BibitemOpen
      \bibfield  {author} {\bibinfo {author} {\bibfnamefont {S.}~\bibnamefont
      {Agarwal}}, \bibinfo {author} {\bibfnamefont {F.~K.}\ \bibnamefont {Chan}},
      \bibinfo {author} {\bibfnamefont {B.}~\bibnamefont {Rallabandi}}, \bibinfo
      {author} {\bibfnamefont {M.}~\bibnamefont {Gazzola}},\ and\ \bibinfo {author}
      {\bibfnamefont {S.}~\bibnamefont {Hilgenfeldt}},\ }\bibfield  {title}
      {\bibinfo {title} {An unrecognized inertial force induced by flow curvature
      in microfluidics},\ }\href@noop {} {\bibfield  {journal} {\bibinfo  {journal}
      {Proc. Natl. Acad. Sci.}\ }\textbf {\bibinfo {volume} {118}},\ \bibinfo
      {pages} {e2103822118} (\bibinfo {year} {2021})}\BibitemShut {NoStop}%
    \bibitem [{\citenamefont {Agarwal}\ \emph {et~al.}(2024)\citenamefont
      {Agarwal}, \citenamefont {Upadhyay}, \citenamefont {Bhosale}, \citenamefont
      {Gazzola},\ and\ \citenamefont {Hilgenfeldt}}]{agarwal2024density}%
      \BibitemOpen
      \bibfield  {author} {\bibinfo {author} {\bibfnamefont {S.}~\bibnamefont
      {Agarwal}}, \bibinfo {author} {\bibfnamefont {G.}~\bibnamefont {Upadhyay}},
      \bibinfo {author} {\bibfnamefont {Y.}~\bibnamefont {Bhosale}}, \bibinfo
      {author} {\bibfnamefont {M.}~\bibnamefont {Gazzola}},\ and\ \bibinfo {author}
      {\bibfnamefont {S.}~\bibnamefont {Hilgenfeldt}},\ }\bibfield  {title}
      {\bibinfo {title} {Density-contrast induced inertial forces on particles in
      oscillatory flows},\ }\href@noop {} {\bibfield  {journal} {\bibinfo
      {journal} {Journal of Fluid Mechanics}\ }\textbf {\bibinfo {volume} {985}},\
      \bibinfo {pages} {A33} (\bibinfo {year} {2024})}\BibitemShut {NoStop}%
    \bibitem [{\citenamefont {Zhang}\ \emph {et~al.}(2024)\citenamefont {Zhang},
      \citenamefont {Minten},\ and\ \citenamefont
      {Rallabandi}}]{zhang2024particle}%
      \BibitemOpen
      \bibfield  {author} {\bibinfo {author} {\bibfnamefont {X.}~\bibnamefont
      {Zhang}}, \bibinfo {author} {\bibfnamefont {J.}~\bibnamefont {Minten}},\ and\
      \bibinfo {author} {\bibfnamefont {B.}~\bibnamefont {Rallabandi}},\ }\bibfield
       {title} {\bibinfo {title} {Particle hydrodynamics in acoustic fields:
      Unifying acoustophoresis with streaming},\ }\href@noop {} {\bibfield
      {journal} {\bibinfo  {journal} {Physical Review Fluids}\ }\textbf {\bibinfo
      {volume} {9}},\ \bibinfo {pages} {044303} (\bibinfo {year}
      {2024})}\BibitemShut {NoStop}%
    \bibitem [{\citenamefont {Doinikov}(1994)}]{doinikov1994acoustic}%
      \BibitemOpen
      \bibfield  {author} {\bibinfo {author} {\bibfnamefont {A.~A.}\ \bibnamefont
      {Doinikov}},\ }\bibfield  {title} {\bibinfo {title} {Acoustic radiation
      pressure on a compressible sphere in a viscous fluid},\ }\href@noop {}
      {\bibfield  {journal} {\bibinfo  {journal} {J. Fluid Mech.}\ }\textbf
      {\bibinfo {volume} {267}},\ \bibinfo {pages} {1} (\bibinfo {year}
      {1994})}\BibitemShut {NoStop}%
    \bibitem [{\citenamefont {Stone}\ and\ \citenamefont
      {Samuel}(1996)}]{stone1996propulsion}%
      \BibitemOpen
      \bibfield  {author} {\bibinfo {author} {\bibfnamefont {H.~A.}\ \bibnamefont
      {Stone}}\ and\ \bibinfo {author} {\bibfnamefont {A.~D.}\ \bibnamefont
      {Samuel}},\ }\bibfield  {title} {\bibinfo {title} {Propulsion of
      microorganisms by surface distortions},\ }\href@noop {} {\bibfield  {journal}
      {\bibinfo  {journal} {Phys. Rev. Lett.}\ }\textbf {\bibinfo {volume} {77}},\
      \bibinfo {pages} {4102} (\bibinfo {year} {1996})}\BibitemShut {NoStop}%
    \bibitem [{\citenamefont {Masoud}\ and\ \citenamefont
      {Stone}(2019)}]{masoud2019reciprocal}%
      \BibitemOpen
      \bibfield  {author} {\bibinfo {author} {\bibfnamefont {H.}~\bibnamefont
      {Masoud}}\ and\ \bibinfo {author} {\bibfnamefont {H.~A.}\ \bibnamefont
      {Stone}},\ }\bibfield  {title} {\bibinfo {title} {The reciprocal theorem in
      fluid dynamics and transport phenomena},\ }\href@noop {} {\bibfield
      {journal} {\bibinfo  {journal} {J. Fluid Mech.}\ }\textbf {\bibinfo {volume}
      {879}} (\bibinfo {year} {2019})}\BibitemShut {NoStop}%
    \bibitem [{\citenamefont {Batchelor}\ and\ \citenamefont
      {Green}(1972)}]{batchelor1972hydrodynamic}%
      \BibitemOpen
      \bibfield  {author} {\bibinfo {author} {\bibfnamefont {G.}~\bibnamefont
      {Batchelor}}\ and\ \bibinfo {author} {\bibfnamefont {J.-T.}\ \bibnamefont
      {Green}},\ }\bibfield  {title} {\bibinfo {title} {The hydrodynamic
      interaction of two small freely-moving spheres in a linear flow field},\
      }\href@noop {} {\bibfield  {journal} {\bibinfo  {journal} {Journal of Fluid
      Mechanics}\ }\textbf {\bibinfo {volume} {56}},\ \bibinfo {pages} {375}
      (\bibinfo {year} {1972})}\BibitemShut {NoStop}%
    \bibitem [{\citenamefont {Brady}\ and\ \citenamefont
      {Bossis}(1988)}]{brady1988stokesian}%
      \BibitemOpen
      \bibfield  {author} {\bibinfo {author} {\bibfnamefont {J.~F.}\ \bibnamefont
      {Brady}}\ and\ \bibinfo {author} {\bibfnamefont {G.}~\bibnamefont {Bossis}},\
      }\bibfield  {title} {\bibinfo {title} {Stokesian dynamics},\ }\href@noop {}
      {\bibfield  {journal} {\bibinfo  {journal} {Annual Review of Fluid
      Mechanics}\ }\textbf {\bibinfo {volume} {20}},\ \bibinfo {pages} {111}
      (\bibinfo {year} {1988})}\BibitemShut {NoStop}%
    \bibitem [{\citenamefont {Durlofsky}\ \emph {et~al.}(1987)\citenamefont
      {Durlofsky}, \citenamefont {Brady},\ and\ \citenamefont
      {Bossis}}]{durlofsky1987dynamic}%
      \BibitemOpen
      \bibfield  {author} {\bibinfo {author} {\bibfnamefont {L.}~\bibnamefont
      {Durlofsky}}, \bibinfo {author} {\bibfnamefont {J.~F.}\ \bibnamefont
      {Brady}},\ and\ \bibinfo {author} {\bibfnamefont {G.}~\bibnamefont
      {Bossis}},\ }\bibfield  {title} {\bibinfo {title} {Dynamic simulation of
      hydrodynamically interacting particles},\ }\href@noop {} {\bibfield
      {journal} {\bibinfo  {journal} {Journal of fluid mechanics}\ }\textbf
      {\bibinfo {volume} {180}},\ \bibinfo {pages} {21} (\bibinfo {year}
      {1987})}\BibitemShut {NoStop}%
    \bibitem [{\citenamefont {Li}\ \emph {et~al.}(2023)\citenamefont {Li},
      \citenamefont {Collis}, \citenamefont {Brumley}, \citenamefont {Schneiders},\
      and\ \citenamefont {Sader}}]{li2023structure}%
      \BibitemOpen
      \bibfield  {author} {\bibinfo {author} {\bibfnamefont {P.}~\bibnamefont
      {Li}}, \bibinfo {author} {\bibfnamefont {J.~F.}\ \bibnamefont {Collis}},
      \bibinfo {author} {\bibfnamefont {D.~R.}\ \bibnamefont {Brumley}}, \bibinfo
      {author} {\bibfnamefont {L.}~\bibnamefont {Schneiders}},\ and\ \bibinfo
      {author} {\bibfnamefont {J.~E.}\ \bibnamefont {Sader}},\ }\bibfield  {title}
      {\bibinfo {title} {Structure of the streaming flow generated by a sphere in a
      fluid undergoing rectilinear oscillation},\ }\href@noop {} {\bibfield
      {journal} {\bibinfo  {journal} {Journal of Fluid Mechanics}\ }\textbf
      {\bibinfo {volume} {974}},\ \bibinfo {pages} {A37} (\bibinfo {year}
      {2023})}\BibitemShut {NoStop}%
    \bibitem [{\citenamefont {Brenner}(1961)}]{bre61_plane}%
      \BibitemOpen
      \bibfield  {author} {\bibinfo {author} {\bibfnamefont {H.}~\bibnamefont
      {Brenner}},\ }\bibfield  {title} {\bibinfo {title} {The slow motion of a
      sphere through a viscous fluid towards a plane surface},\ }\href@noop {}
      {\bibfield  {journal} {\bibinfo  {journal} {Chemical engineering science}\
      }\textbf {\bibinfo {volume} {16}},\ \bibinfo {pages} {242} (\bibinfo {year}
      {1961})}\BibitemShut {NoStop}%
    \bibitem [{\citenamefont {Rallabandi}\ \emph {et~al.}(2017)\citenamefont
      {Rallabandi}, \citenamefont {Hilgenfeldt},\ and\ \citenamefont
      {Stone}}]{ral17_hydroforce}%
      \BibitemOpen
      \bibfield  {author} {\bibinfo {author} {\bibfnamefont {B.}~\bibnamefont
      {Rallabandi}}, \bibinfo {author} {\bibfnamefont {S.}~\bibnamefont
      {Hilgenfeldt}},\ and\ \bibinfo {author} {\bibfnamefont {H.~A.}\ \bibnamefont
      {Stone}},\ }\bibfield  {title} {\bibinfo {title} {Hydrodynamic force on a
      sphere normal to an obstacle due to a non-uniform flow},\ }\href@noop {}
      {\bibfield  {journal} {\bibinfo  {journal} {Journal of Fluid Mechanics}\
      }\textbf {\bibinfo {volume} {818}},\ \bibinfo {pages} {407} (\bibinfo {year}
      {2017})}\BibitemShut {NoStop}%
    \end{thebibliography}

%

\end{document}